\documentclass[review]{elsarticle}

\usepackage{lineno,hyperref}
\usepackage{graphicx}
\usepackage{epstopdf}
\usepackage{algorithm}
\usepackage{amsmath}  
\usepackage{algorithmicx}
\usepackage{algpseudocode} 
\usepackage{booktabs}
\usepackage{multirow}
\usepackage[utf8]{inputenc}
\usepackage{url}
\usepackage{amssymb}
\usepackage{bbding}
\usepackage{wasysym}
\usepackage{utfsym}
\usepackage{fontawesome}
\usepackage{float}
\usepackage{array}
\usepackage{subcaption}

\modulolinenumbers[20]

\begin{document}
\begin{frontmatter}
\title{A large-scale distributed parallel discrete event simulation engines based on Warped2 for Wargaming simulation}
\author[a]{Xiaoning Jia\fnref{myfootnote1}}
\author[a]{Ruilin Kong\fnref{myfootnote1}}
\author[c]{Guangya Si}
\author[c]{Bilong Shen\corref{mycorrespondingauthor}}\ead{blshen@email.cn}
\author[a,b]{Zhe Ji\corref{mycorrespondingauthor}}\ead{jizhe@nwpu.edu.cn}
\address[a]{School of Software, Northwestern Polytechnical University, Xi'an, Shanxi, 710129, China}
\address[b]{Yangtze River Delta Research Institute of NPU Northwestern Polytechnical University, Taicang, Jiangsu, 215400, China}
\address[c]{College of Joint Operation, National Defense University, Beijing 100091, China}
\fntext[myfootnote1]{These authors contributed equally to this work.}
\cortext[mycorrespondingauthor]{Corresponding authors.}

\begin{abstract}
    Rising demand for complex simulations highlights conventional engines' scalability limits, spurring Parallel Discrete Event Simulation (PDES) adoption. 
    Warped2, a PDES engine leveraging Time Warp synchronization with Pending Event Set optimization, delivers strong performance, it struggles with inherent wargaming limitations: inefficient LP resource allocation during synchronization and unaddressed complex entity interaction patterns.
    To address these challenges, we present an optimized framework featuring four synergistic improvements: (1) Asynchronous listener threads are introduced to address event monitoring latency in large-scale scenarios, instead of synchronous polling mechanisms, (2) METIS-based load rebalancing strategy is incorporated to address the issue of dynamic event allocation during real-world simulation, (3) Entity interaction solver with constraint satisfaction mechanisms is designed to mitigate state conflicts, and (4) Spatial hashing algorithm to overcome O(n²) complexity bottlenecks in large-scale nearest-neighbor searches. 
    Experimental validation through a GridWorld demo demonstrates significant enhancements in temporal fidelity and computational efficiency. 
    Benchmark results show our framework achieves $16\times$ acceleration over baseline implementations and maintains $8\times$ speedup over 1-thread configuration across MPI and Pthreads implementations. 
    The combined load balancing and LP migration strategy reduces synchronization overhead by 58.18\%, with load balancing accounting for 57\% of the total improvement as the dominant optimization factor. 
    These improvements provide an enhanced solution for PDES implementation in large-scale simulation scenarios.
\end{abstract}

\begin{keyword}
\texttt Parallel Discrete Event Simulation \sep Warped2 \sep Engine Architecture \sep Wargaming Simulation
\end{keyword}

\end{frontmatter}

\section{Introduction}
    In recent decades, Wargaming Simulation has become an essential tool for decision support, military training, and strategic planning\cite{Li2021}. 
    These simulations leverage large-scale simulations to model complex interactions in dynamic and uncertain environments\cite{Riebl2021}. 
    As the scale and complexity of simulation scenarios increase, traditional engines face significant challenges in efficiently managing large numbers of interacting entities while maintaining scalability and real-time performance\cite{Leong2020}. 
    Ensuring accuracy and consistency in large-scale discrete-event simulations, particularly in high-throughput combat and dynamic strategic settings, remains a critical challenge\cite{Barr2023}. 
    Overcoming these limitations requires advancements in simulation methodologies, particularly through the integration of parallel computing\cite{Cavalcante2020}.

    Wargaming Simulation plays a vital role across various domains. 
    As modern warfare increasingly integrates multiple domains—land, sea, air, space, and cyber—Wargaming Simulation has become indispensable for exploring new operational models\cite{Gonzalez2024}.
    Large-scale simulations face significant challenges, mainly due to computational complexity and data synchronization\cite{Johnson2024}. These simulations, involving numerous entities and complex interactions, place heavy demands on computational resources, which increase exponentially as system size grows\cite{Lin2023}. Traditional algorithms struggle to handle such workloads efficiently. Moreover, ensuring data synchronization and consistency is difficult when multiple participants collaborate within the same virtual environment, where real-time data transmission is essential for maintaining simulation accuracy and integrity.
    Recent research has focused on key areas, including AI-driven Wargaming Simulation, which leverages deep learning and reinforcement learning to optimize tactical decisions, inspired by models like AlphaGo\cite{Li2021}. 
    Large-scale distributed simulations, utilizing cloud computing, high-performance computing (HPC), and federation architectures (e.g., HLA), enhance computational capacity\cite{Cavalcante2020}. Multi-agent simulations (MAS) employ agent-based models to simulate unit interactions, improving realism. Additionally, data-driven modeling, combining big data and machine learning, aims to boost simulation accuracy and adaptability\cite{Gonzalez2024}.
    
    The rapid advancement of parallel computing has significantly influenced PDES, enabling efficient handling of large-scale, complex simulations\cite{Barr2023}. 
    Traditional sequential discrete-event simulators struggle with performance bottlenecks in computationally intensive applications such as wargaming and large-scale network simulations\cite{Leong2020}. 
    PDES addresses these challenges by leveraging multi-core and distributed architectures to parallelize event execution while minimizing synchronization overhead\cite{Lin2023}.

    A key breakthrough in PDES is the Time Warp mechanism, a speculative execution strategy allowing LPs to advance independently and roll back if causality violations occur\cite{Johnson2024}. Several PDES frameworks have integrated Time Warp and other optimizations to improve scalability and efficiency. Among them, Warped2 stands out as a configurable open-source simulation kernel featuring optimized Time Warp execution and an efficient Pending Event Set mechanism, making it well-suited for wargaming and large-scale simulations\cite{Wilsey2019, Gupta2018}.
    Beyond Warped2, other PDES frameworks have introduced advancements in parallel performance, including:

    \begin{itemize}
        \item \textbf{ROSS}(Rensselaer's Optimistic Simulation System): Optimized for HPC clusters with efficient event management and load balancing.
        \item \textbf{Simian}: Uses Just-in-Time (JIT) compilation to accelerate large-scale distributed simulations.
        \item \textbf{µsik}: Focuses on real-time performance and fault tolerance, making it suitable for defense and cybersecurity simulations.
        \item \textbf{ROOT-Sim}(The ROme OpTimistic Simulator): Implements an optimized Time Warp mechanism with efficient rollback and incremental state management, enhancing scalability for large-scale parallel simulations.
    \end{itemize}
    
    Despite these advances, applying existing PDES frameworks directly to wargame simulation presents challenges, including the need for domain-specific event processing, real-time decision-making, and fine-grained resource management. Additionally, fully exploiting modern hardware architectures such as GPUs remains an open research area.
    And although the Warped2 kernel has been optimized for complex model architecture and interface design, it still exhibits several shortcomings that affect its application for war game engines.
    
    \textbf{Internal shortcomings} include:
    \begin{itemize}
        \item Inadequate mechanisms for handling events in large-scale throughput.
        \item Inability to autonomously adjust the load of each process.
    \end{itemize}

    \textbf{External shortcomings} include:
    \begin{itemize}
        \item Lack of interfaces for entity interaction, crucial for war game simulations.
    \end{itemize}

    For large-scale distributed simulations of Wargaming, it is essential to redesign the engine architecture to address these issues and incorporate parallelism\-based entity interaction algorithms.
    In response to the aforementioned shortcomings, this paper presents the following node-based improvements and extensions to the Warped2 framework, specifically targeting large-scale wargaming simulations.

    \begin{enumerate}[1)]       
        \item A distributed Warped framework for dynamic event scheduling and high-throughput simulation system modeling, supporting flexible simulation modifications and additional inputs. The garbage collection mechanism in the original framework was revised. The thread coordination mechanism of the original Warped2 kernel was optimized by introducing listener threads, enhancing the system's robustness and integrity.
        \item A load balancing module for the distributed framework. The modified module maps thread IDs to scheduling queue IDs, allowing threads to retrieve and process events from different scheduling queues, thereby balancing thread workloads. The METIS library built into Warped2 is further optimized for load balancing, significantly improving system performance.
        \item The reconstruction of the logical process and event communication structure of the simulation model based on Warped2 achieves effective strategy simulation model design and large-scale parallelism. Through simulation correctness and model performance tests, large-scale simulation performance is enhanced by an average of 16 times.
        \item A nearest-neighbor entity search algorithm based on a mesh improves the overall computational speed of the solver through techniques such as mesh cube coordinate system offsets and storing mesh hash tables. This approach enhances the performance of nearest-neighbor search by an average of 34 times compared to brute-force search methods.
    \end{enumerate}

    The rest of the paper is organized as follows. 
    Section 2 details the structure and mechanisms of Warped2, justifying its selection from various engine architectures. 
    Section 3 focuses on addressing the internal shortcomings of Warped2, specifically optimizing the engine for large-scale, complex scenarios.
    Section 4 tackles the external shortcomings of Warped2 by designing a solver interface for large-scale entity interaction. 
    Section 5 validates the optimized framework through experiments on load balancing, correctness, system and algorithm performance, and demonstrates its practical potential with a GridWorld demo.

\section{Related Works}

    There are several popular implementations of Time Warp synchronized parallel simulation engines. 
    In this section, we first provides a detailed introduction to the architecture and mechanisms of the Warped2 parallel simulation engine, focusing on its time-warping mechanism, event scheduling strategies and parallel processing model.
    It then examines the architecture and implementation of existing mainstream parallel discrete event simulation frameworks, followed by a comparative analysis with Warped2. 
\subsection{Architecture and Mechanism of Warped2}
    
\begin{figure}[H]   
        \centering
        \includegraphics[width=\textwidth]{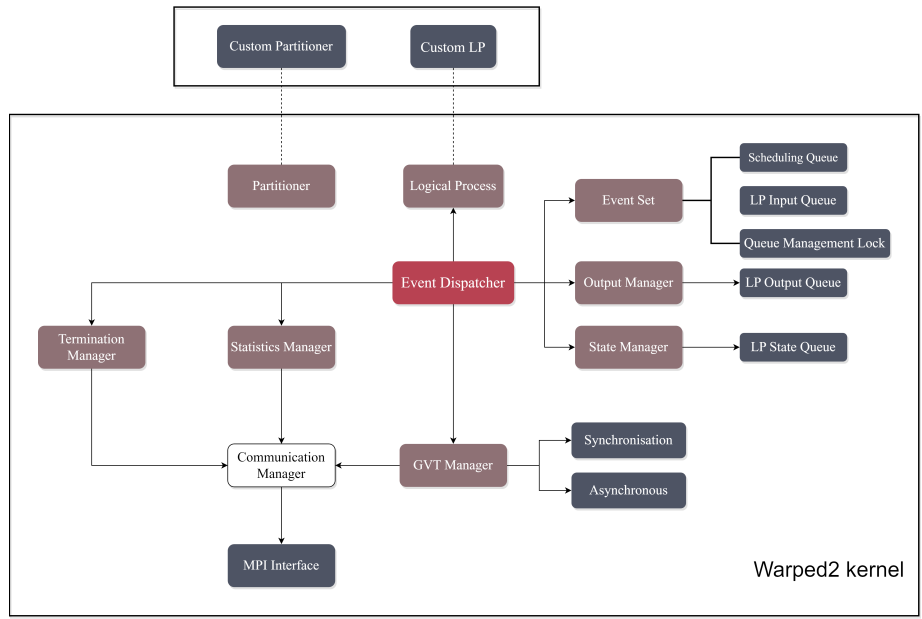}
        \caption{Warped2 Kernel}
        \label{img1}
    \end{figure}

    Wilsey et al.\cite{Wilsey2019} propose an open-source Time Warp simulation kernel with high configurability and offers various time warp optimization features for exploration called Warped2, the kernel is composed of five major components: the Event Dispatcher, the Global Manager, the Communication Manager, the Statistics Manager, and the Termination Manager, as illustrated in Figure \ref{img1}, the Warped2 kernel architecture.
    
    These Time Warp components operate in concert, utilizing MPI messages for inter-node communication, which are divided into local and global categories. 
    Local components are responsible for managing specific activities of all logical processes on the current computing node, which include the Event Dispatcher and its subcomponents. 
    Global components are responsible for overseeing and executing the computational tasks of the entire parallel simulation cluster, including the Global Virtual Time Manager, the Statistics Manager, the Termination Manager, and their subcomponents\cite{Gupta2018}. 

\begin{figure}[htbp]
        \centering
        \includegraphics[width=\textwidth]{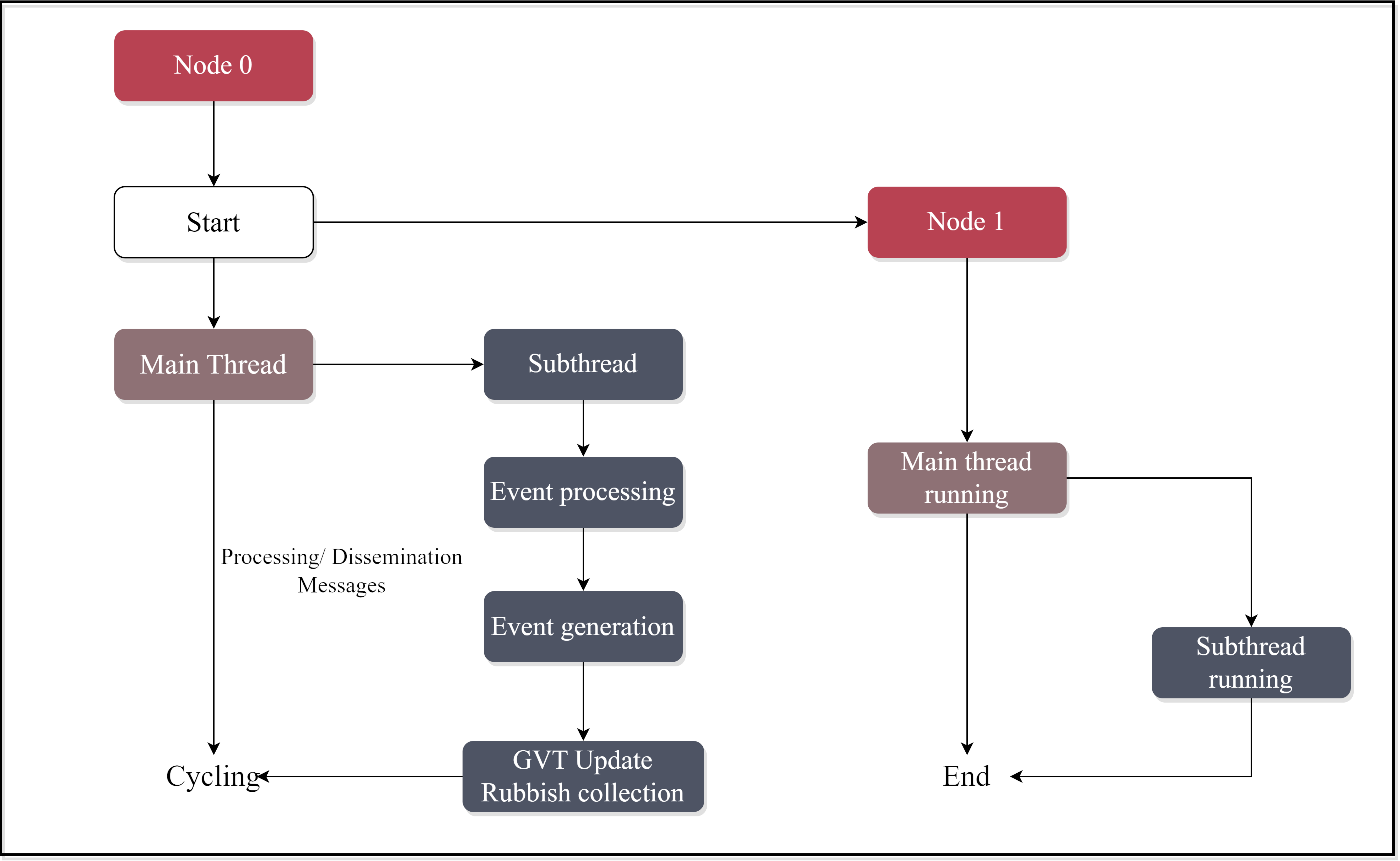}
        \caption{Warped2 Workflow}
        \label{img2}
    \end{figure}

    In the local components, the Event Dispatcher include managing the event set, maintaining the state of simulation units and implementing garbage collection, which involves event rollback mechanisms. 
    The event set subcomponent primarily consists of data structures for handling all pending and processed events on the current computing node, including the scheduling queue, LP input queue and management locks. 
    The output manager subcomponent is used to store and track all output queues required for the current node's event outputs in Warped2. The state manager contains a data structure for an LP state queue, which stores all saved LP states.
    In the global components, there are three main parts: the Global Virtual Time Manager, the Statistics Manager, and the Termination Manager. 
    The Global Virtual Time Manager is responsible for monitoring the global progress of the parallel simulation process. 
    The Termination Manager oversees the progress of processes during the simulation runtime, while the Statistics Manager records local node simulation data and provides methods for the kernel to compute and organize comprehensive statistical data for the entire simulation, which is then presented to researchers.
    Through the coordination of local and global components, Warped2, as proposed by Wilsey et al., supports an autonomous simulation workflow once the initial conditions are determined, as illustrated in Figure \ref{img2}. 
    
    In Warped2, each computing node has both a main thread and a worker thread. 
    The main thread handles overall message communication for the node, while the worker thread manages the processing of events. 
    Each node maintains its own state and manages its worker threads; Once event processing is complete, a series of messaging mechanisms is employed to manage the state of threads and processes within a node. 
    In the Warped2 workflow, each event is assigned a fully ordered clock value based on virtual time. 
    Warped2 employs a straightforward event processing loop for event handling, as outlined in Algorithm \ref{alg1}.
    
    However, our research has identified that while Warped2 possesses certain strengths, it still has limitations and vulnerabilities in terms of system scalability, complexity, performance, and real-time capabilities. 
    The original Warped framework was effective for execution on single-core clusters, but its kernel architecture was not well-suited for parallelization and inter-process communication. 
    Therefore, to meet our needs, we have undertaken optimizations based on Warped2, focusing on system scalability and complexity, performance and real-time capabilities, and parallel event simulation.
    
    \begin{algorithm}[htbp]
        \caption{Processing loop for event handling}
        \label{alg1}
        {\bf Require:}\\ 
        \hspace*{0.05in} Input: $scheduled event$\\
        \hspace*{0.05in} Output: $The LP state is modified or the queue is updated$\\
        \begin{algorithmic}[1] 
            \While{not detected signal to terminate}
                \State $e = getNextScheduledEvent()$
                \State $LP = receiver of e$
                \If {e < last processed event for LP}
                    \State $rollback(LP)$
                    \If {isAntiMessage(e)}
                        \State $cancelEvent(e) (If possible)$
                        \State $scheduleNewEvents(LP)$
                        \State $continue$
                    \EndIf
                \EndIf
                \State $processEvent(e)$
                \State $saveState(LP)$
                \State $sendNewlyGeneratedEvents()$
                \State $moveToProcessedQueue(e)$
                \State $replaceScheduledEvent(LP)$
            \EndWhile
        \end{algorithmic}
    \end{algorithm}
    
\subsection{The ROme OpTimistic Simulator (ROOT-Sim)}
    ROOT-Sim (ROme OpTimistic Simulator)\cite{Pellegrini2011}is a Time Warp-based parallel discrete event simulation framework using MPI for large-scale simulations. 
    It supports dynamic thread allocation, event scheduling, rollback, and memory optimization via Dynamic Memory Logger and Restorer (DyMeLoR). 
    ROOT-Sim also features a Committed and Consistent Global Snapshot (CCGS) service for reconstructing global snapshots after each Global Virtual Time (GVT) calculation.

    In large-scale parallel discrete event simulations, Warped2 is often preferred over ROOT-Sim due to its superior performance in handling event scheduling, rollback mechanisms and memory management. 
    While ROOT-Sim offers features like dynamic thread allocation and memory optimization\cite{Pellegrini2011}, these add complexity and may lead to inefficiencies in large, multi-core environments. 
    In contrast, Warped2 provides a simpler, more efficient approach to event processing, with better fine-grained rollback control\cite{Gupta2018}, which enhances scalability in large systems. 
    Additionally, Warped2 excels in global virtual time (GVT) computation\cite{Barr2023}, offering more reliable and efficient handling of event dependencies\cite{Johnson2024}.

\subsection{Rensselaer’s Optimistic Simulation System (ROSS)}
    ROSS (Runtime Optimistic Simulation System)\cite{Carothers2000} is a Time Warp-based parallel discrete event simulation framework designed for large-scale simulations. 
    It uses MPI for communication and supports dynamic event scheduling, rollback mechanisms, and parallel GVT calculation. 
    ROSS is optimized for high performance and scalability, making it suitable for large, complex simulations across distributed systems.

    In large-scale parallel discrete event simulations, Warped2 is preferred over ROSS due to its simpler memory management\cite{Bucci2022} and more efficient event scheduling\cite{Warped2}. 
    While ROSS offers optimized time-stepping and parallel GVT calculations\cite{Fujimoto2016}, its complexity can lead to performance bottlenecks in high-concurrency environments. 
    Warped2, on the other hand, provides better scalability and stability\cite{Warped2}, making it more suitable for large-scale distributed simulations.

\subsection{Current Simulation-Based Gaming Systems and Limitations of Existing Warped2 Models}
    Contemporary simulation-based gaming systems leverage centralized inference architectures (e.g., SEED RL’s TPU clusters\cite{Espeholt2020}) and hierarchical reinforcement learning (HRL\cite{Kulkarni2016}) to address large-scale state-space complexity and multi-agent coordination, achieving 11× training acceleration via Actor-Learner parallelization\cite{Espeholt2020} and human-AI hybrid strategies (e.g., StarCraft II Centaur Mode). 
    While these systems excel in tactical decision-making, challenges persist in exploration-exploitation balance and heterogeneous environment generalization. 
    
    In contrast, existing Warped2 Models exhibits domain-specific limitations: traffic models lack dynamic congestion adaptation, epidemic simulations rely on static transmission parameters, and PCS models incur high concurrency costs from frequent base station handovers. 
    Systemic constraints—including ARM incompatibility, rigid schedulers, and inefficient temporal backtracking—further hinder scalability. 
    Bridging these gaps necessitates adaptive scheduling algorithms and hybrid partitioning strategies to optimize causal consistency and hardware utilization, aligning Warped2 with modern simulation demands for real-time responsiveness and cross-domain flexibility.

\section{Kernel Optimization of Warped2}
    In this section, we analyze two key issues in the Warped2 core framework—insufficient event handling in large-scale throughput and the lack of autonomous load balancing across processes. 
    To address these, we propose optimizations using listener threads and METIS-based partitioning, laying the foundation for the development of the entity interaction solver. 
    
    \subsection{Large-scale throughput Event Handling: Listener threads}
    To address the high-throughput and large-scale load requirements of simulation systems, we propose further development of the distributed Warped framework to meet these needs. 
    In earlier sections, we thoroughly discussed the messaging and processing mechanisms of Warped2. Warped2, as a discrete event simulation framework, only supports autonomous execution after the initial conditions are set, without the flexibility to modify the simulation or introduce additional input during runtime. 
    This limitation renders it inadequate for our purposes, requiring architectural changes. Specifically, when a simulation completes, there should be a mechanism to allow the injection of new events to continue the simulation. 
    Additionally, during an ongoing simulation, there must be an interruption mechanism to pause execution and await user input for adjustments before resuming.
    
    The initialization logic for each simulation unit should focus on translating user commands into corresponding events and dispatching them. Furthermore, the garbage collection mechanism requires modification. 
    Currently, it collects all events with timestamps earlier than the Global Virtual Time (GVT) at each GVT update, leading to frequent collection operations. 
    To optimize this, we propose collecting events after a fixed number of GVT iterations, retaining more events to ensure that processed events can be rolled back If needed, especially after new user input, thus preventing premature garbage collection.
    Each node maintains its own node state and manages its own worker threads. In practice, after event processing is completed, there is a series of messaging mechanisms in place to manage the thread and process state of a node.

    \begin{figure}[htbp]
        \centering
        \includegraphics[width=\textwidth]{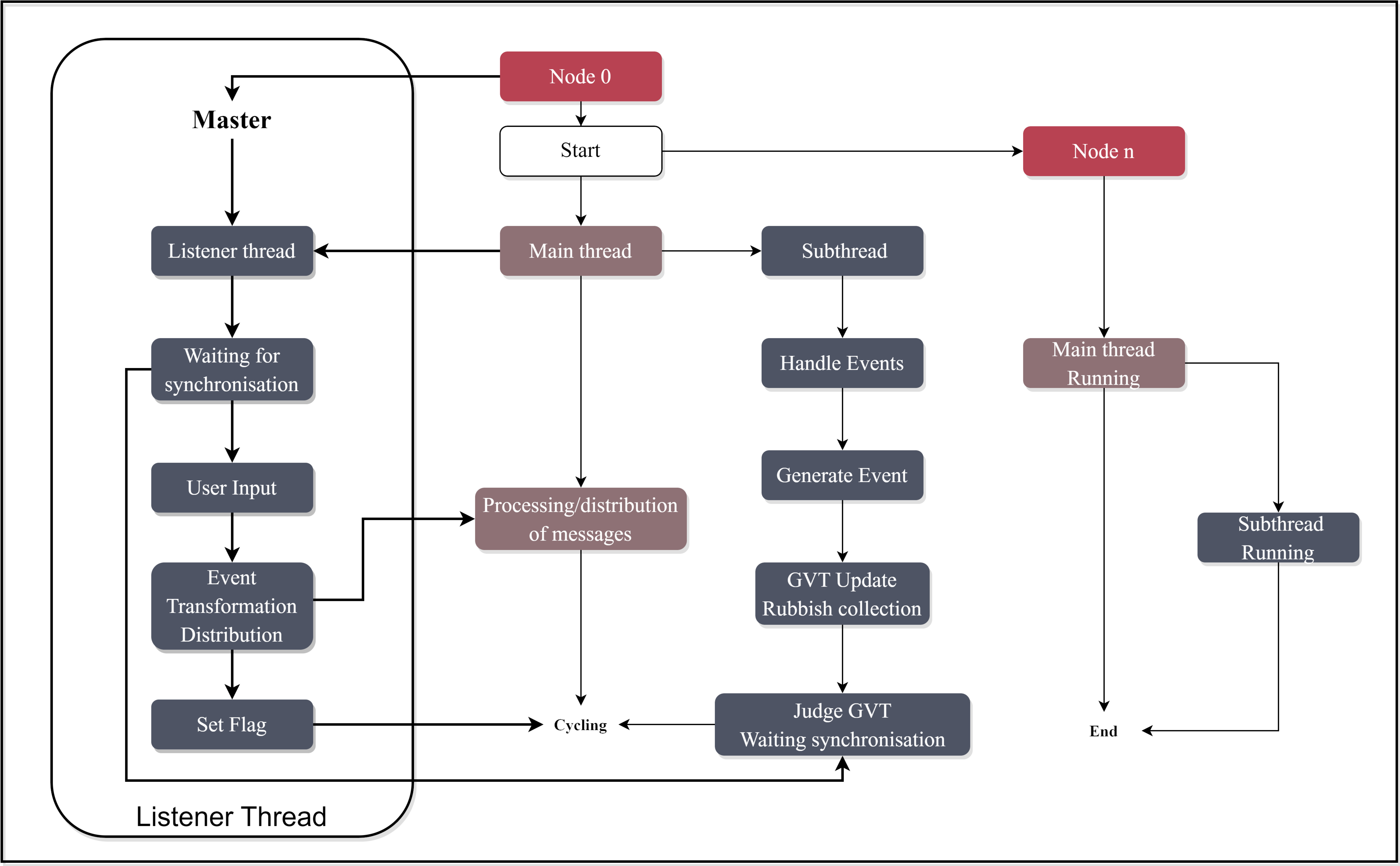}
        \caption{Warped2 Optimised Process}
        \label{img3}
    \end{figure}

    Figure \ref{img2} illustrates the original thread coordination mechanism of the Warped2 kernel. 
    To implement the proposed improvements, the architecture needs to be modified, and Figure \ref{img3} presents a flowchart comparison of the new process. 
    Notably, a new thread has been introduced to manage the high-throughput coordination mechanism in addition to the main thread and worker threads. 
    This listening thread waits for additional user input, ensuring the system's high-throughput feasibility. 
    The introduction of a dynamic event addition mechanism allows for real-time control over the simulation's direction as it progresses. 
    In the creation of the listening thread, additional operations were performed on the main node, adding a flag to coordinate the program's execution. 
    When user input is received, the termination manager prevents the simulation from ending. 
    Simultaneously, an MPI messaging interface notifies other nodes that the listening thread is active and additional events will be input. 
    At this point, If all threads have completed processing their events, they will enter an idle state, awaiting either new event input or program termination instructions.

\subsection{Load balancing for Node Event Processing: METIS-based re-partitioning}
    In the simulation, each unit processes events and generates new ones, which may necessitate adjustments to the load-balancing strategy initially configured during program setup. 
    Ensuring balanced computational workloads across different nodes during runtime is critical. 
    In earlier test cases, the initial distribution of events and simulation units was relatively uniform, leading to consistent workloads across nodes and reducing the need for significant load-balancing efforts.
    
    However, in real-world simulations, different simulation units represent entities with varying data volumes and interaction frequencies. 
    Consequently, load balancing becomes crucial.
    The Warped2 parallel discrete-event simulation framework employs pre-simulation analysis to construct a weighted graph model characterizing event communication patterns among LPs. 
    A multi-level k-way static partitioning optimization is implemented using the METIS graph partitioning library, following a three-phase process: 
    \begin{itemize}
        \item[1.] Graph coarsening through vertex merging
        \item[2.] Minimum-cut partitioning on the reduced graph
        \item[3.] Boundary refinement through reverse expansion
    \end{itemize}

    \begin{figure}[htbp]
        \centering
        \includegraphics[width=\textwidth]{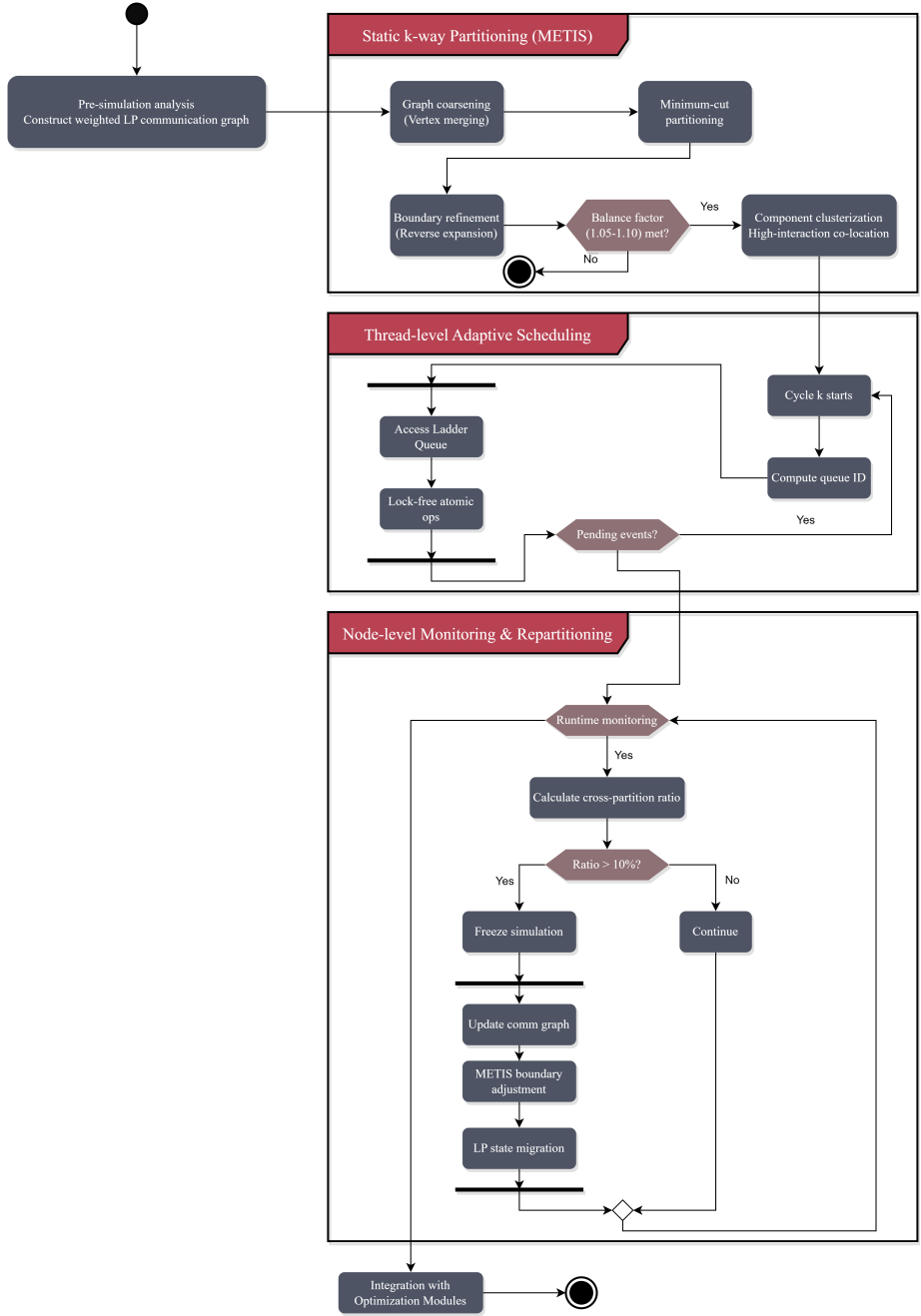}
        \caption{METIS-based Adaptive Repartitioning Mechanism Flow}
        \label{tag:rePartitioning Mechanism}
    \end{figure}
    This partitioning strategy incorporates a balance factor (1.05-1.10) to constrain LP quantity disparities between partitions while minimizing cross-partition communication volumes. 
    Optimized partitions maintain static mapping through component clusterization, co-locating highly interactive elements (e.g., RAID controllers with associated disk arrays in storage models).

    To address dynamic load fluctuations, a dual\-layer adaptive mechanism is implemented as shown in Figure \ref{tag:rePartitioning Mechanism}. At the thread level, dynamic scheduling employs periodic modulo-N thread ID hashing (e.g., thread $T_i$ accessing queue $Q_{(i+k)modN}$ at cycle k) combined with lock\-free atomic operations in Ladder Queue structures. 
    Globally, a threshold-triggered incremental repartitioning mechanism (activated when cross-partition events exceed 10\%) initiates simulation suspension, communication graph updates, METIS boundary adjustments, and LP state migration. 
    Through tight integration with communication managers (optimizing MPI/shared-memory routing), rollback managers (intra-partition state snapshots), and event schedulers, this architecture achieves significant reductions in communication overhead and synchronization latency while preserving temporal causality constraints in parallel discrete-event simulations. 

    The mechanism works by associating each thread with a scheduling queue based on its thread ID. 
    When a thread retrieves and processes events, it operates on its corresponding queue. 
    To enhance load balancing, an additional step is introduced: the mapping between thread IDs and scheduling queue IDs is periodically recalculated. 
    This allows threads to pull events from different queues in each cycle, effectively distributing the workload evenly among all threads.

\section{Wargaming Simulation Solver for Complex Entity Interaction}

    In this Section, building upon the optimized Warped2 core framework from Section 3, we develop the entity interaction interface within the Warped2 framework. 
    We design a Wargaming Simulation Solver based on complex entity interactions, modify the LP design, and optimize the entity interaction algorithms for specific scenarios.
\subsection{Redesigning LP Definitions for Wargaming Simulations}
    In Warped2, Wilsey et al. propose three simulation models: Epidemic Disease Propagation, Portable Cellular Service (PCS), and Traffic, where grids are defined as LPs and entity actions as events. 
    While this approach works for certain scenarios, it does not meet the needs of wargaming simulations that require detailed analysis of entity parameters and interactions. 
    We have revised the event definitions, using continuous entity actions and real-world warfare interactions as the basis for event creation, and developed a solver where entities are LPs and their actions are events.

    \subsection{Grid for Given Rules-Based Model}

    We require a solver model based on specific simulation rules, with a redefinition of the model grid.
	To ensure the accuracy and effectiveness of both the simulation rules and entity movement strategies, we drew inspiration from real-world warfare scenarios, extracting action rules for entities and events from established wargame simulation frameworks and military strategy models. Our approach incorporates principles from classic hex-and-counter systems, such as PanzerBlitz and Advanced Squad Leader (ASL), which utilize turn-based movement, action points (AP), and zone of control (ZOC) mechanisms to simulate tactical engagements. 
    To enhance the realism of command structures and decision-making processes, we take reference from John Boyd’s OODA loop (Observe-Orient-Decide-Act), a widely recognized model for rapid tactical adaptation. Furthermore, modern digital wargame systems such as Command: Modern Operations inform our approach to simultaneous-action mechanics (WEGO).
    
    After extensive comparison, these established wargame methodologies were selected as the foundation for entity rules and simulation strategy design due to their proven effectiveness in military simulations and strategy game development. Their widespread application and high fidelity in system simulation ensure that our model achieves a balance between accuracy and realism.

    After defining the specific simulation rules, the design of the background grid and its coordinate system becomes necessary to implement entity simulation. 
    We referenced the H3 hexagonal grid spatial indexing system, abstracting the background grid into a hexagon-based Multi-agent Gridworld. 
    The overall Gridworld is designed with a single-tier structure, where each grid cell has a unique hash index based on its spatial location, enabling efficient position data lookup and processing. 
    Given that the background grid consists of regular hexagons, there are currently four available coordinate system designs. Table 1 compares these four coordinate systems from the perspectives of storage, computation, and symmetry.

    \begin{table}[H]
        \centering
        \caption{Coordinate System Comparison}
        \label{tab:Coordinate System Comparison}
        \resizebox{\textwidth}{!}{
            \begin{tabular}{ccccc}
                \toprule
                Coordinate Name & Misalignment & Multiplication & Vertically & Cubic \\
                \midrule
                Vector Operation & 	$\times$ & \checkmark & \checkmark & \checkmark\\
                Hash Storage & \multicolumn{4}{c}{Arbitrary Shape} \\
                Array Storage & Square & Hard & Dimensional & Hard \\
                Hexagonal Symmetry & $\times$ & \checkmark & \checkmark & \checkmark \\
                Basic Arithmetic & Less Frequent & Less & Most Members & All \\
                \bottomrule
            \end{tabular}
        }
    \end{table}

\begin{figure}[htbp]
        \centering
        \includegraphics[width=0.75\textwidth]{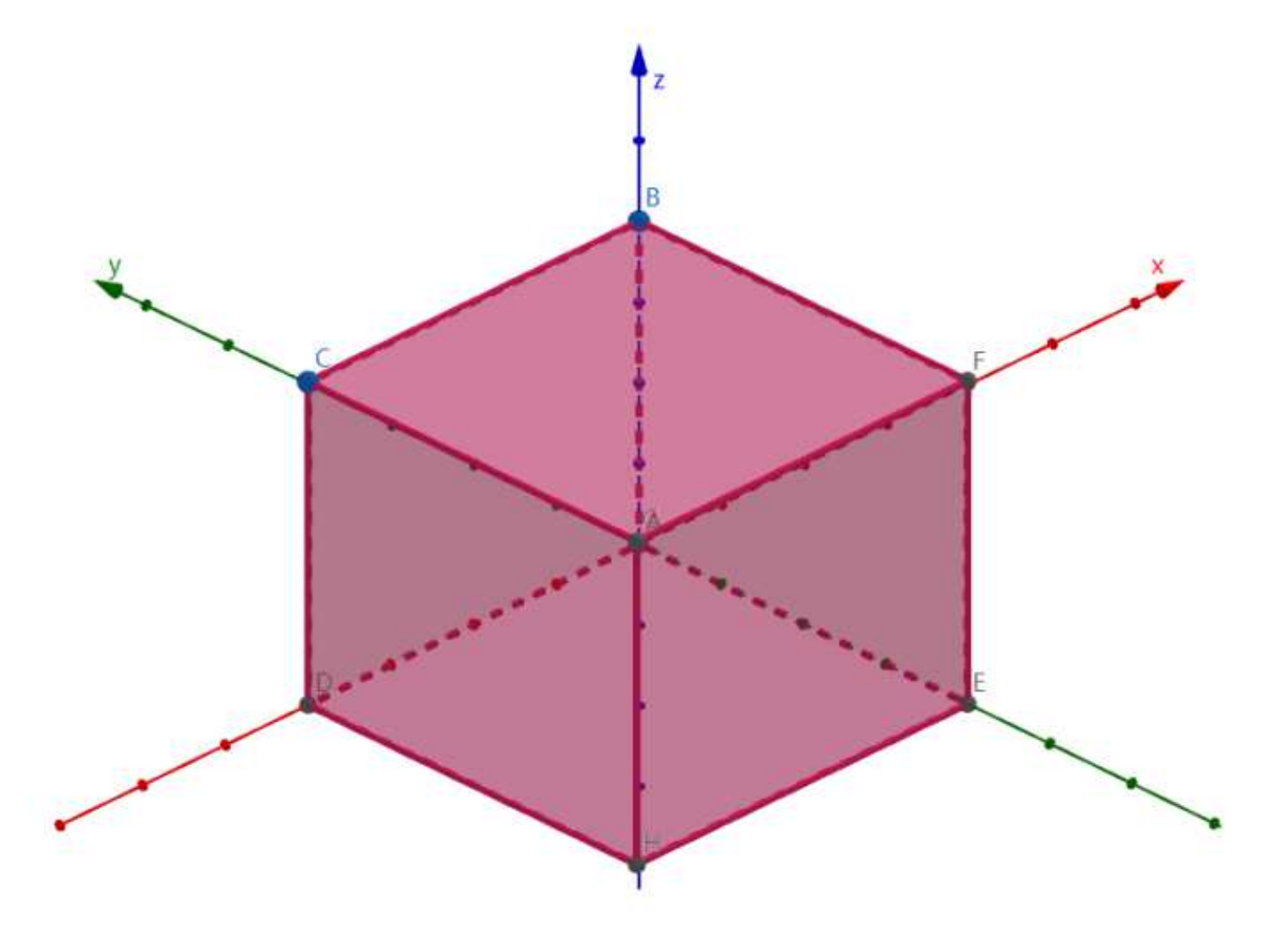}
        \caption{Cube Coordinate}
        \label{tab:Cube Coordinate}
    \end{figure}

    Table \ref{tab:Coordinate System Comparison} clearly demonstrates that the cubic coordinate system has distinct advantages in terms of computation, hash storage and symmetry. 
    In terms of array storage, the cubic coordinate system can be converted into offset coordinates by keeping one coordinate axis constant. 
    Therefore, we opted for the cubic coordinate system, which utilizes three primary directional axes and exhibits strong symmetry, as shown in Figure \ref{tab:Cube Coordinate}.

    \begin{figure}[htbp]
        \centering
        \includegraphics[scale=0.5]{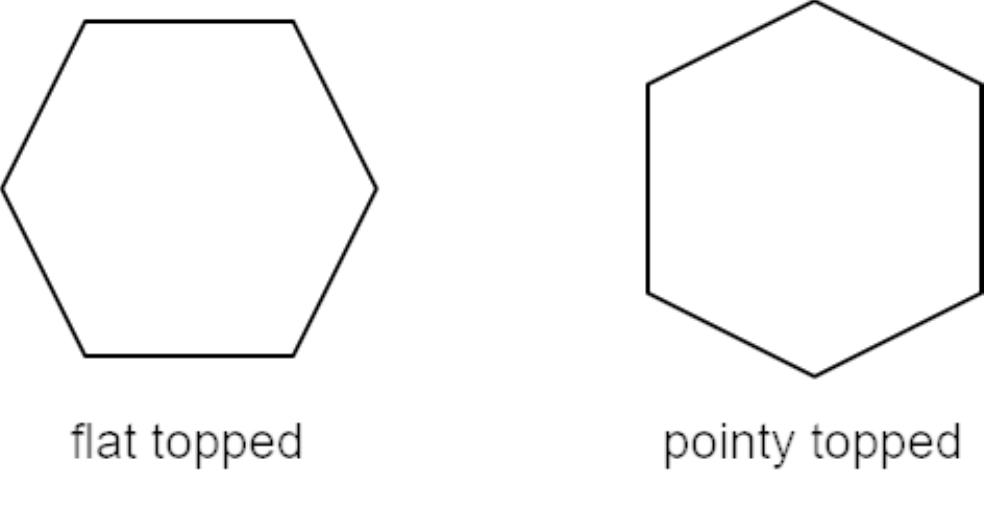}
        \caption{Regular Hexagonal Grid Formation}
        \label{tab:Regular Hexagonal Grid Formation}
    \end{figure}

    \begin{figure}[htbp]
        \centering
        \includegraphics[width=0.75\textwidth]{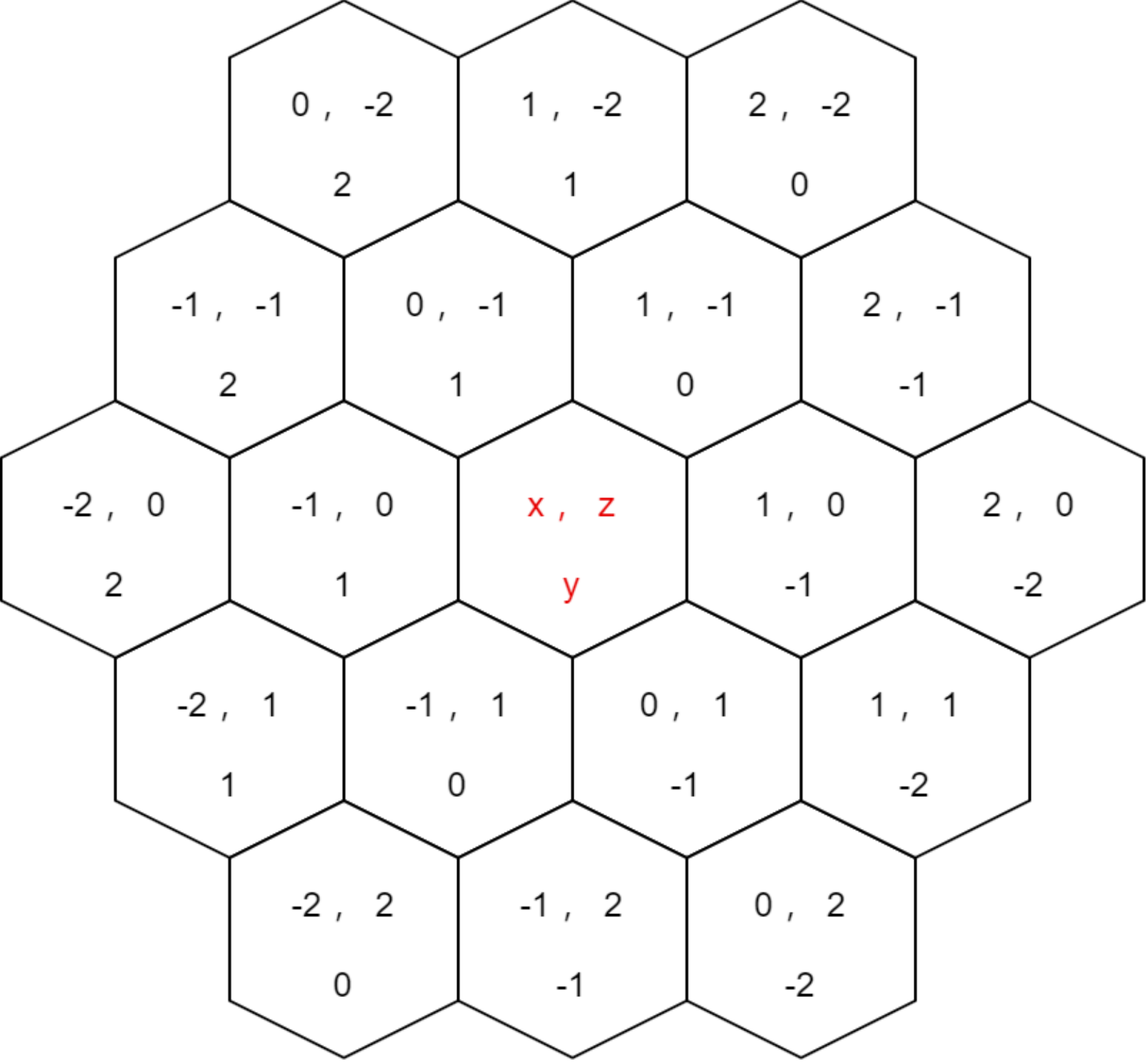}
        \caption{Cube Coordinates Layout}
        \label{tab:Cube Coordinates Layout}
    \end{figure}

    By selecting a diagonal plane within the cube that satisfies the coordinate constraint x+y+z=0,we can obtain a simple regular hexagonal cube. 
    The regular hexagon itself has two forms, as shown in Figure \ref{tab:Regular Hexagonal Grid Formation}. 
    This configuration is beneficial for subsequent mathematical calculations and map construction. In the header file, we designed an enumeration variable entity\_direction\_t to store the labels of the six directions of the hexagon, corresponding to the six edges of the hexagonal grid. 
    According to the distinction of north, south, east, and west in Figure \ref{tab:Cube Coordinates Layout}, NORTH\_EAST corresponds to the northern half\-axis where Y remains unchanged, EAST corresponds to the eastern half\-axis where Z remains unchanged, SOUTH\_EAST corresponds to the southern half\-axis where X remains unchanged, SOUTH\_WEST corresponds to the southern half\-axis where Y remains unchanged, WEST corresponds to the western half\-axis where Z remains unchanged, and NORTH\_WEST corresponds to the northern half\-axis where X remains unchanged.

    After redefining the model grid, the next crucial task is to establish a representative simulation rule. 
    Drawing on Mozi as our reference source, we carefully analyzed the real-world data on entities and events from the Mozi database. 
    By selectively extracting relevant data and applying custom simulation rules, we convert real-world entities into abstract entities and data to ensure the accuracy of the simulation.
    
\begin{figure}[H]
        \centering
        \includegraphics[width=\textwidth]{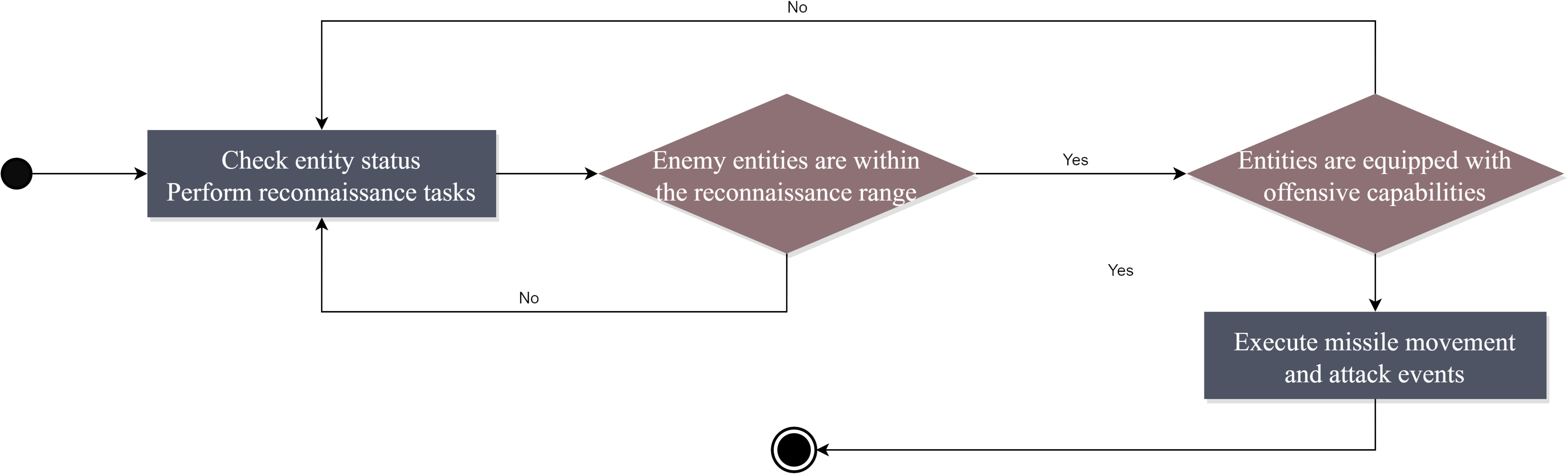}
        \caption{Entity Workflow}
        \label{tab:Entity Workflow}
    \end{figure}

    In our analysis, we found that the detection and attack ranges of real-world entities are influenced by the equipment they carry. 
    To streamline our solver design, we set fixed detection and attack ranges for each entity type, ensuring that entities of the same type have uniform capabilities. 
    This allows us to focus on efficient event handling, scheduling, and parallel performance optimization.
    Given the need for entities with varying data scales, we scripted the generation of these entities. 
    Entities were categorized into four types: ground buildings, aircraft, ground forces, and surface vessels. 
    For each category, we selected a representative entity, with all entities of the same type assigned identical detection and attack ranges, as detailed in Table \ref{tab:Selection of Real-world Case}.
    In combination with the permanent entity cases from Table \ref{tab:Selection of Real-world Case}, the operational workflow of the entities can be summarized and illustrated as shown in Figure \ref{tab:Entity Workflow}.

    \begin{table}[H]
        \centering
        \caption{Selection of Real-world Case}
        \label{tab:Selection of Real-world Case}
        \resizebox{\textwidth}{!}{
            \begin{tabular}{ccccc}
                \toprule
                Name & Type & Detection Range & Attack Range & Movement Speed \\
                \midrule
                Airport NO.5 & 2x4000m+runways & / & / & / \\
                F-16CM & Fighter Aircraft & 110km & 110km & 888.96km/h \\
                T-14 Armata & Tank & 65km & 65km & 120.6km/h \\
                Jinjiang Missile Corvette & Warship & 115km & 115km & 55.56km/h \\
                \bottomrule
            \end{tabular}
        }
    \end{table}

    We have inductively designed four types of permanent entities based on a cubic coordinate system and the context of a regular hexagonal grid, as presented in Table \ref{tab:Unified Fixed Attributes of Entities}. 
    Additionally, we developed a Python script for random entity generation based on these properties, which can produce a specified number of distinct entity types with varying cubic coordinates within a defined grid range.

    \begin{table}[H]
        \centering
        \caption{Unified Fixed Attributes of Entities}
        \label{tab:Unified Fixed Attributes of Entities}
        \resizebox{\textwidth}{!}{
            \begin{tabular}{cccc}
                \toprule
                Type Name & Detection Range & Attack Range & Movement Speed \\
                \midrule
                Ground Structures & / & / & / \\
                Aircraft & 2 grid & 2 grid & 4 grid / interval \\
                Ground Forces & 1 grid & 1 grid & 2 grid / interval \\
                Vessels & 2 grid & 2 grid & 2 grid / interval \\
                \bottomrule
            \end{tabular}
        }
    \end{table}

\subsection{Solver Structure}
\begin{figure}[htbp]
        \centering
        \includegraphics[width=\textwidth]{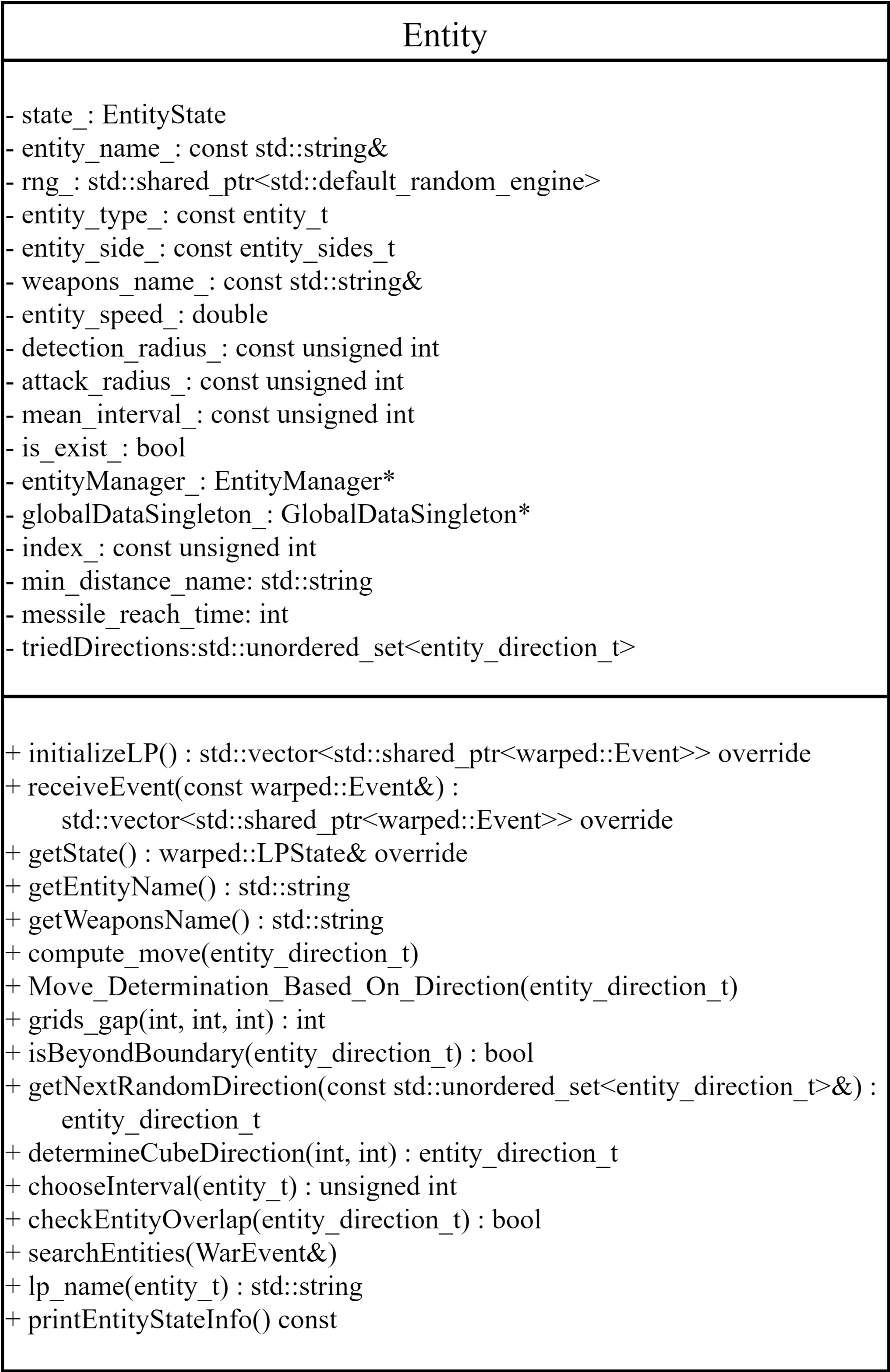}
        \caption{LP Structure}
        \label{tab:LP Structure}
    \end{figure}

    For the design of the simulation solver architecture, we have divided it into three main modules: custom LP structure, event structure and global manager structure. 
    In the original simulation model library of Warped2, models such as epidemics and volcanic eruptions define the grid itself as LP, while entities are defined as events and event handlers. 
    This approach does not meet our design requirements. Therefore, in the custom LP design module, we define the entity itself as an LP, which is an entity class that inherits from the LogicalProcess class in Warped2. 
    The specific structure is shown in Figure \ref{tab:LP Structure}.

    In our simulation framework, entities are modeled as Logical Processes (LPs) that participate in the simulation. 
    Before the simulation begins, the initial dataset for these entities is configured via command-line arguments using TCLAP's ValueArg class, which loads the dataset as an Arg type to create a Simulation object in Warped2. 
    Each LP has two key methods: $initializeLP()$, which handles initialization, and receiveEvent(const warped::Event \&event), which processes valid events (excluding anti-messages and stragglers) and sends them to the Warped2 event scheduler once time synchronization is confirmed.

    The LP class includes frequently updated attributes like entity state, along with static properties such as faction, speed, weapon type, detection range, and attack range. 
    It also holds pointers to two global managers: entityManager and globalDataSingleton. The entity state (state\_) is defined as a structure using the WARPED\_DEFINE\_LP\_State\_STRUCT macro, which outlines the mutable attributes of the state during the simulation.
    This design of the entity state structure eliminates the need for manually writing lengthy inheritance and method definition code.
    
\begin{figure}[htbp]
        \centering
        \includegraphics[scale=0.16]{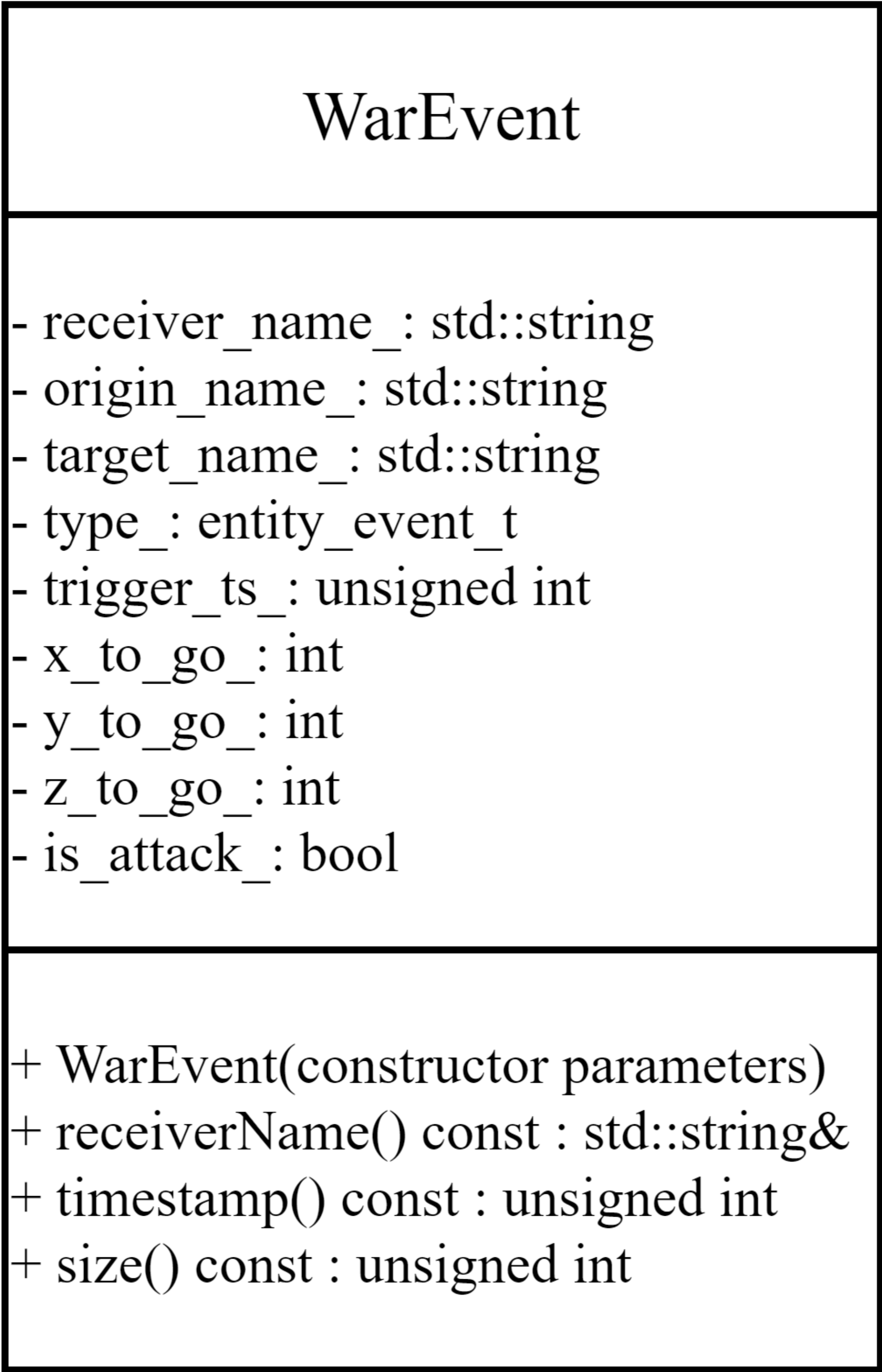}
        \caption{Event Structure}
        \label{tab:Event Structure}
    \end{figure}

    Upon finalizing the custom LP structure design ,which means establishing complete entity, attention must be given to entity movement, specifically the design of the event structure within the solver. 
    Based on our research, Warped2 processes LPs by sequentially handling events in three separate queues associated with the current LP. 
    Following this principle, we developed the event structure shown in Figure \ref{tab:Event Structure}.

    \begin{figure}[htbp]
        \centering
        \includegraphics[width=\textwidth]{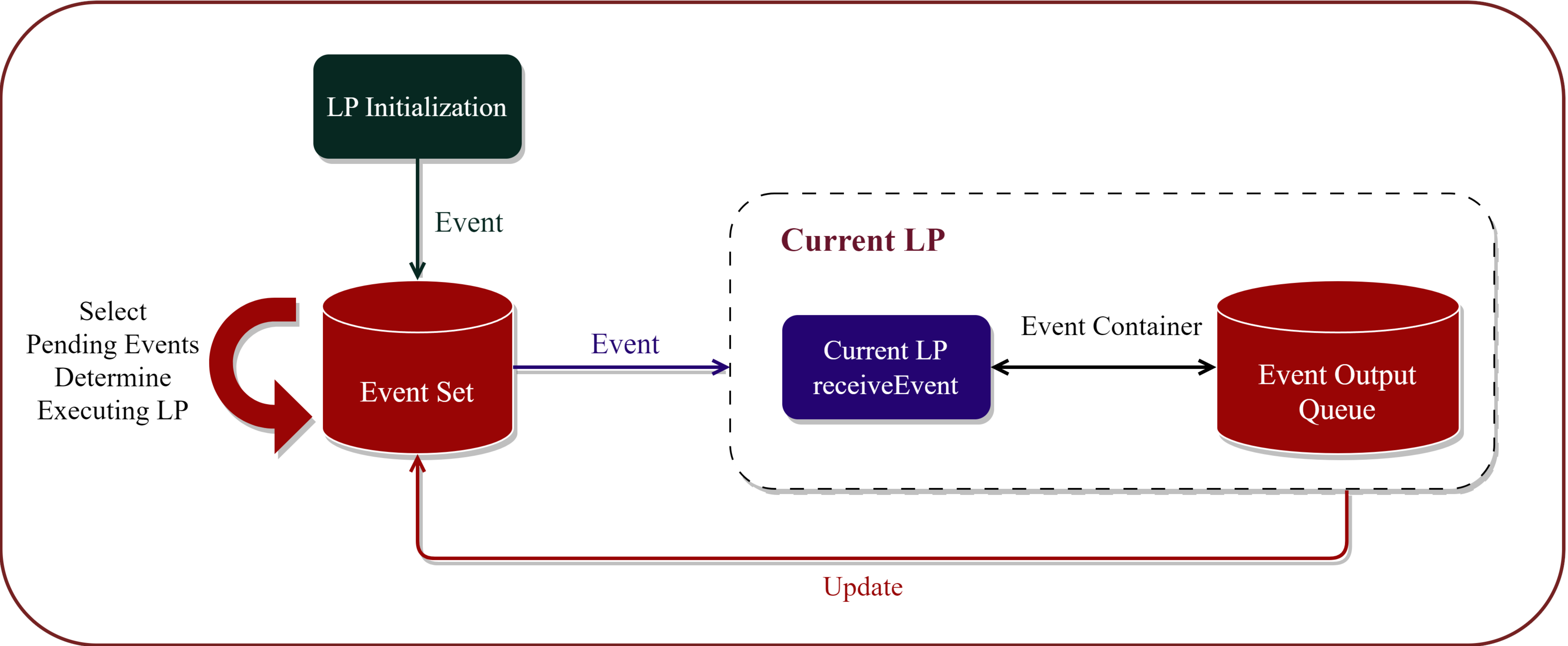}
        \caption{Event Workflow}
        \label{tab:Event Workflow}
    \end{figure}

    The WarEvent, an extension of Warped2's Event, is divided into three types: reconnaissance, attack, and missile movement events. 
    During simulation, the reconnaissance event is the first to execute upon LP initialization, updating the global entity state and potentially generating other events. 
    The event flow is shown in Figure \ref{tab:Event Workflow}.
    Events created at LP initialization are stored in a pending queue and processed in order, with the target LP determined by the WarEvent's recipient. 
    The simulation starts with reconnaissance events, which may trigger missile movement events If a target is found. 
    Upon impact, an attack event destroys the target entity, continuing the cycle of reconnaissance, missile movement, and destruction.
    To ensure persistence, event data is serialized using WARPED\_REGI\allowbreak STER\_SERIALIZABLE\_MEMBERS, enabling efficient storage and cross\-platform transmission in distributed simulations.

\begin{figure}[htbp]
        \centering
        \includegraphics[width=\textwidth]{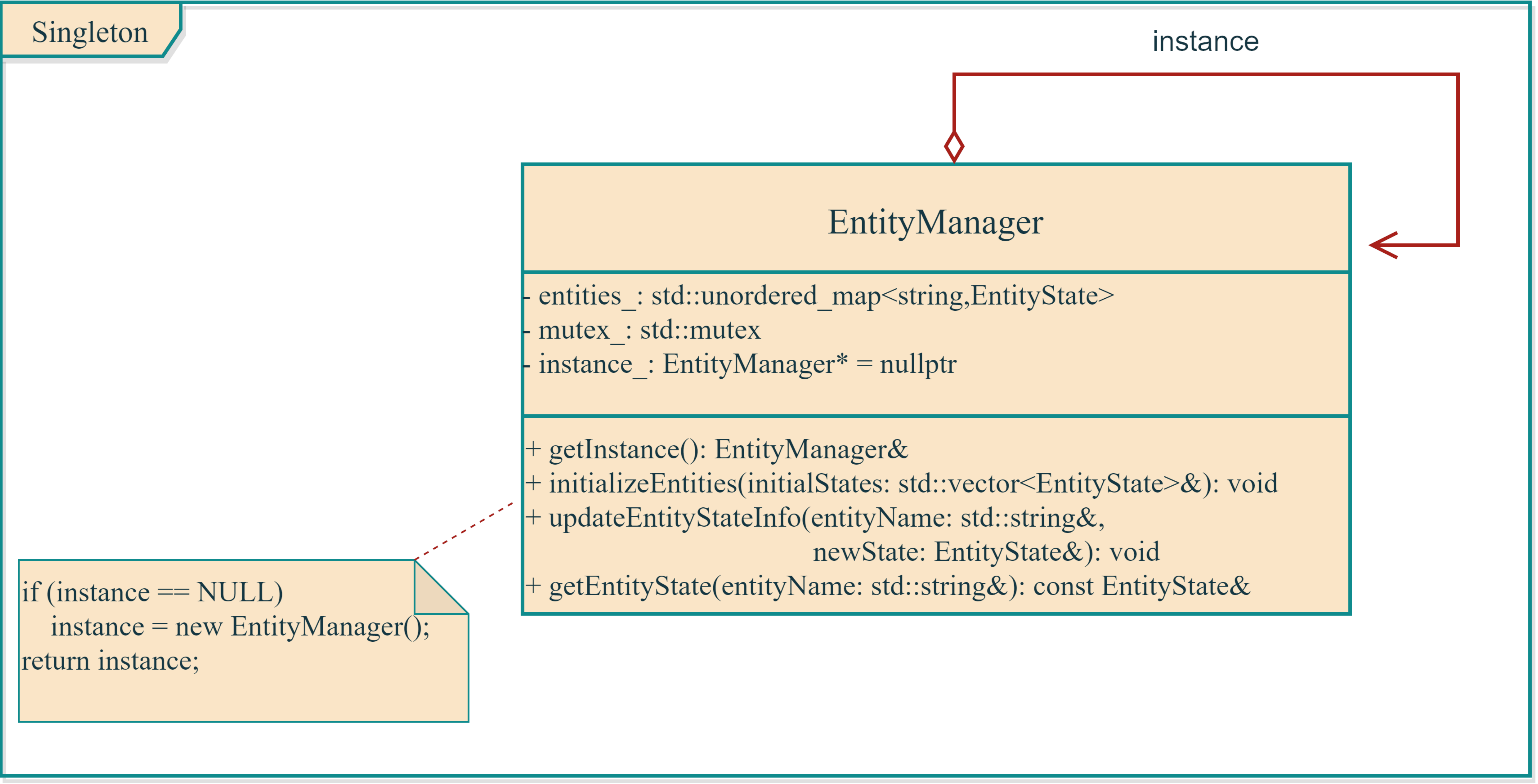}
        \caption{Global LP State Manager}
        \label{tab:Global LP State Manager}
    \end{figure}

    Upon completing the design of the first two components, the solver framework began to take form. 
    However, it became evident that substantial interaction and information exchange between LPs during simulation necessitated a data-sharing mechanism, leading to the design of the global manager structure.
    This shared data structure comprises two components: a global manager for LP states pertaining to entities. During the processing of events through the receiveEvent function, an LP lacks access to the real-time states of other LPs. 
    The global manager enables real-time mutex updates and ensures uniqueness, constructed using the singleton pattern. 
    The EntityManager functions as a global access point for this purpose, as illustrated in Figure \ref{tab:Global LP State Manager}.
    The solver initializes the global state manager for the initial entity data through the initializeEntities method. 
    During the LP's event processing, each update to the entity state is explicitly synchronized with the global state manager using the updateEntityStateInfo method, thereby fulfilling the requirement for real-time global sharing.

\begin{figure}[htbp]
        \centering
        \includegraphics[width=\textwidth]{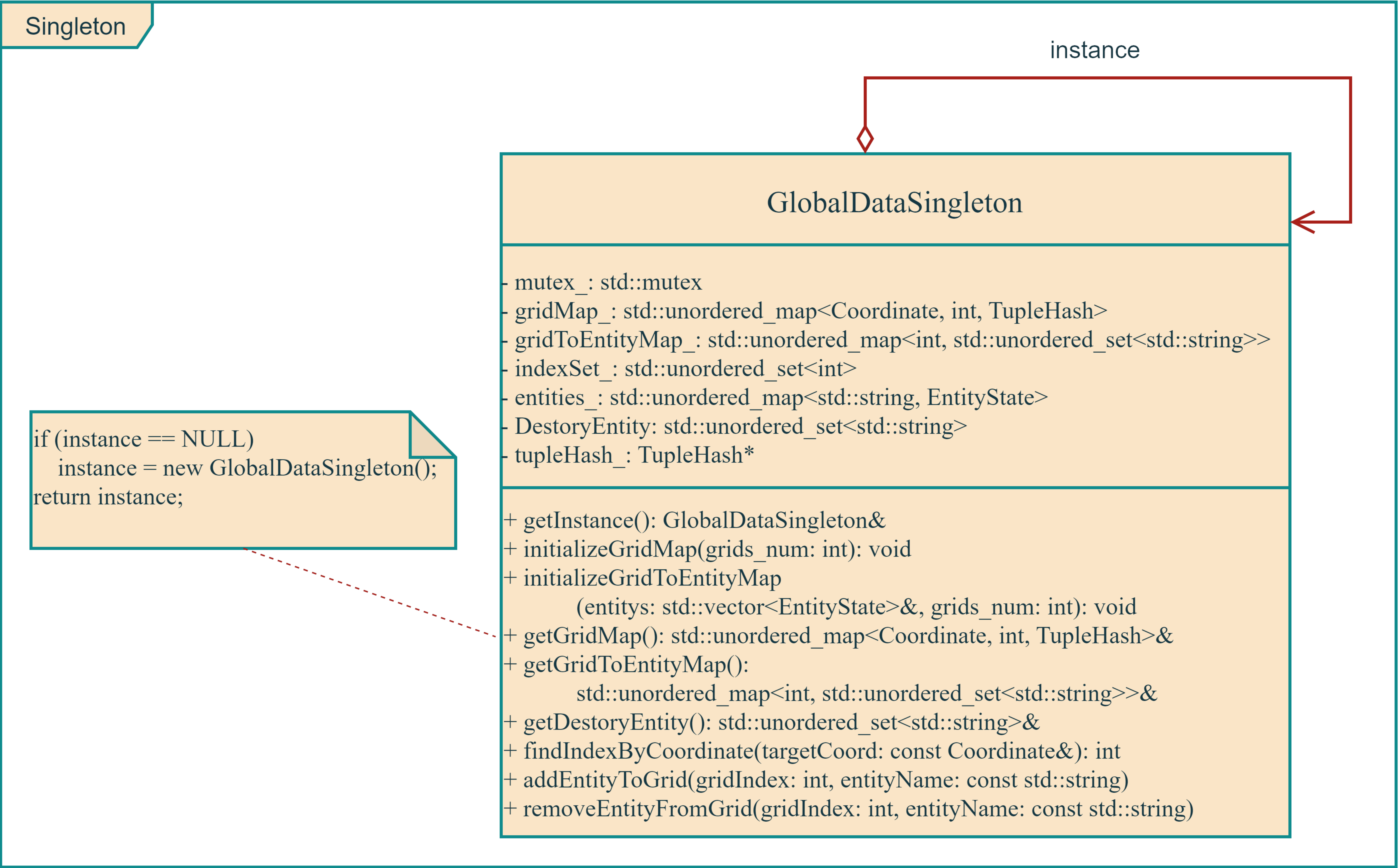}
        \caption{Global Hash Table Manager}
        \label{tab:Global Hash Table Manager}
    \end{figure}

    The second component of shared data focuses on the global information management of the hash table. 
    The solver requires real-time updates and global sharing of entity data for each grid during the calculation of neighboring entities, as detailed in Section \ref{section:Entity Search in Event Processing}. 
    To facilitate this, we designed the hash table to be incorporated within the same singleton global manager, as illustrated in Figure \ref{tab:Global Hash Table Manager} which presents a clear depiction of the hash table's global state manager structure. 
    The mutex\_ serves as a thread-safe mutex, while gridMap\_ is an unordered map that primarily manages the mapping of grid coordinates to indices. 
    The gridToEntityMap\_ is utilized to store mappings from grids to lists of entities; due to the uniqueness of entity names, they can serve as identifiers, thereby reducing unnecessary memory overhead. 
    The singleton pattern guarantees the uniqueness of these critical hash tables while enabling synchronized real-time modifications of gridToEntityMap\_.

\subsection{Event Processing}
\begin{figure}[htbp]
        \centering
        \includegraphics[width=\textwidth]{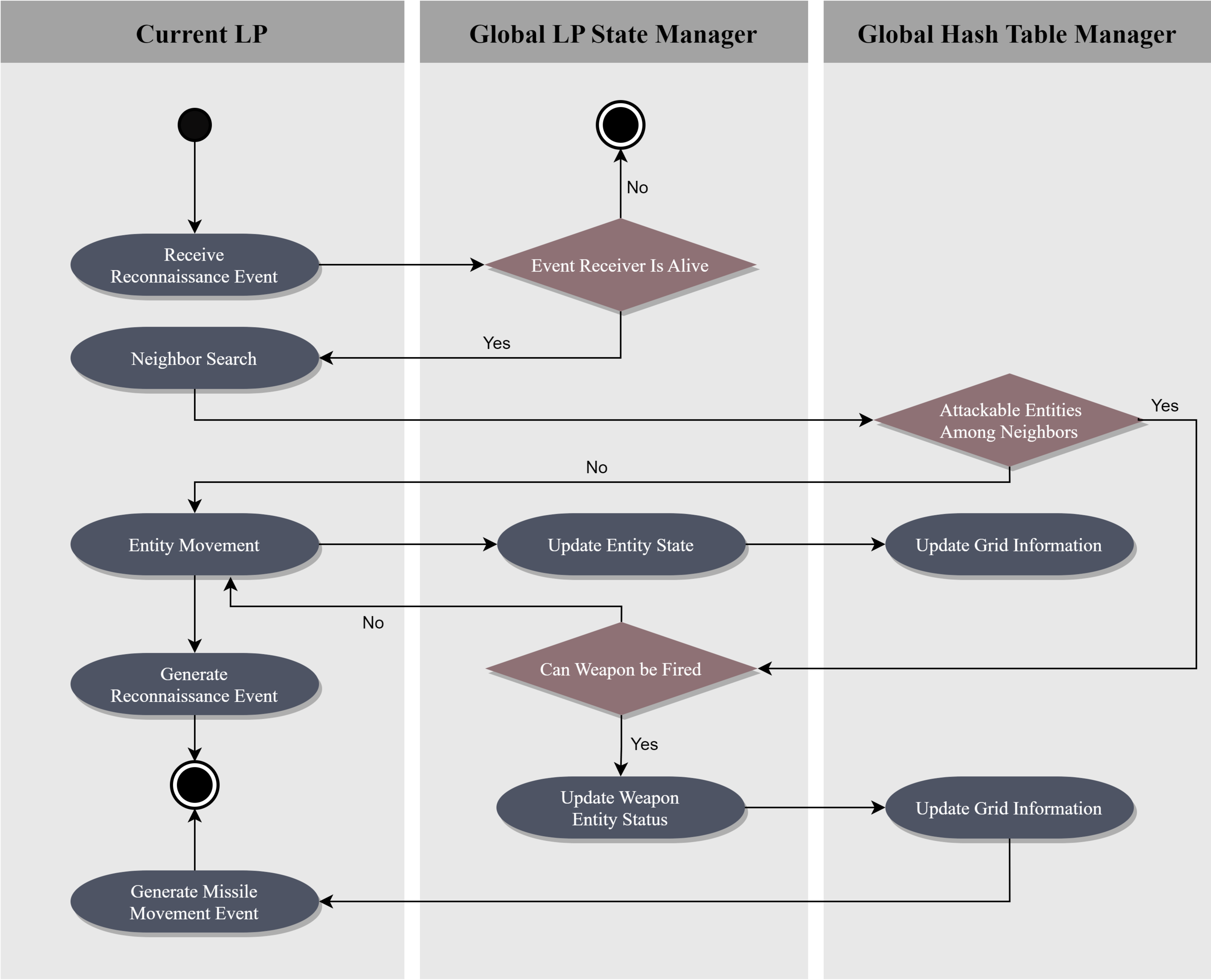}
        \caption{Logic of Reconnaissance Events}
        \label{tab:Logic of Reconnaissance Events}
    \end{figure}

    After finalizing the LP structure, event structure, and global manager, the solver's design is complete. 
    The next step is to ensure correct simulation execution, focusing on the interaction logic between events, which are divided into three types: reconnaissance, attack, and missile movement. 
    Reconnaissance events are the foundation, initialized for all LPs, involving event generation, global manager updates, and the neighbor search algorithm from Section \ref{section:Entity Search in Event Processing}. 
    When an LP receives a reconnaissance event, the global manager checks If the LP still exists. If it does, a neighbor search is conducted via the grid hash manager. 
    If no attackable entities are found, the LP moves randomly, updates its state, and generates a new reconnaissance event. Otherwise, a missile movement event is triggered, as shown in Figure \ref{tab:Logic of Reconnaissance Events}.

\begin{figure}[htbp]
        \centering
        \includegraphics[width=\textwidth]{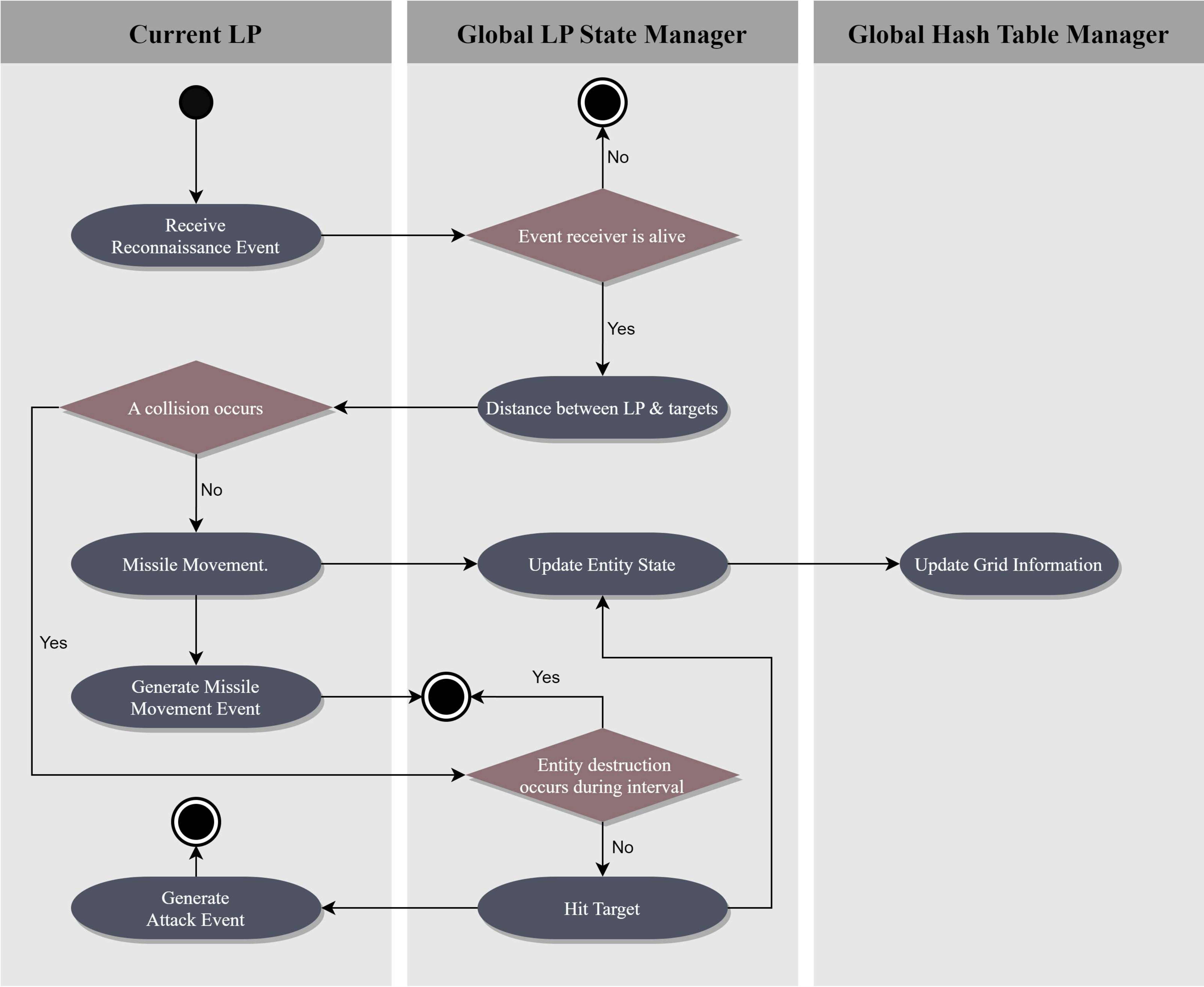}
        \caption{Logic of Missile Movement Events}
        \label{tab:Logic of Missile Movement Events}
    \end{figure}

    The missile movement event is triggered when an LP detects an attackable enemy during a reconnaissance event and its weapon is ready. 
    The event, with the LP as $origin\_name\_$, moves and attacks based on the missile's state and attributes, as shown in Figure \ref{tab:Logic of Missile Movement Events}.

\begin{figure}[htbp]
        \centering
        \includegraphics[width=\textwidth]{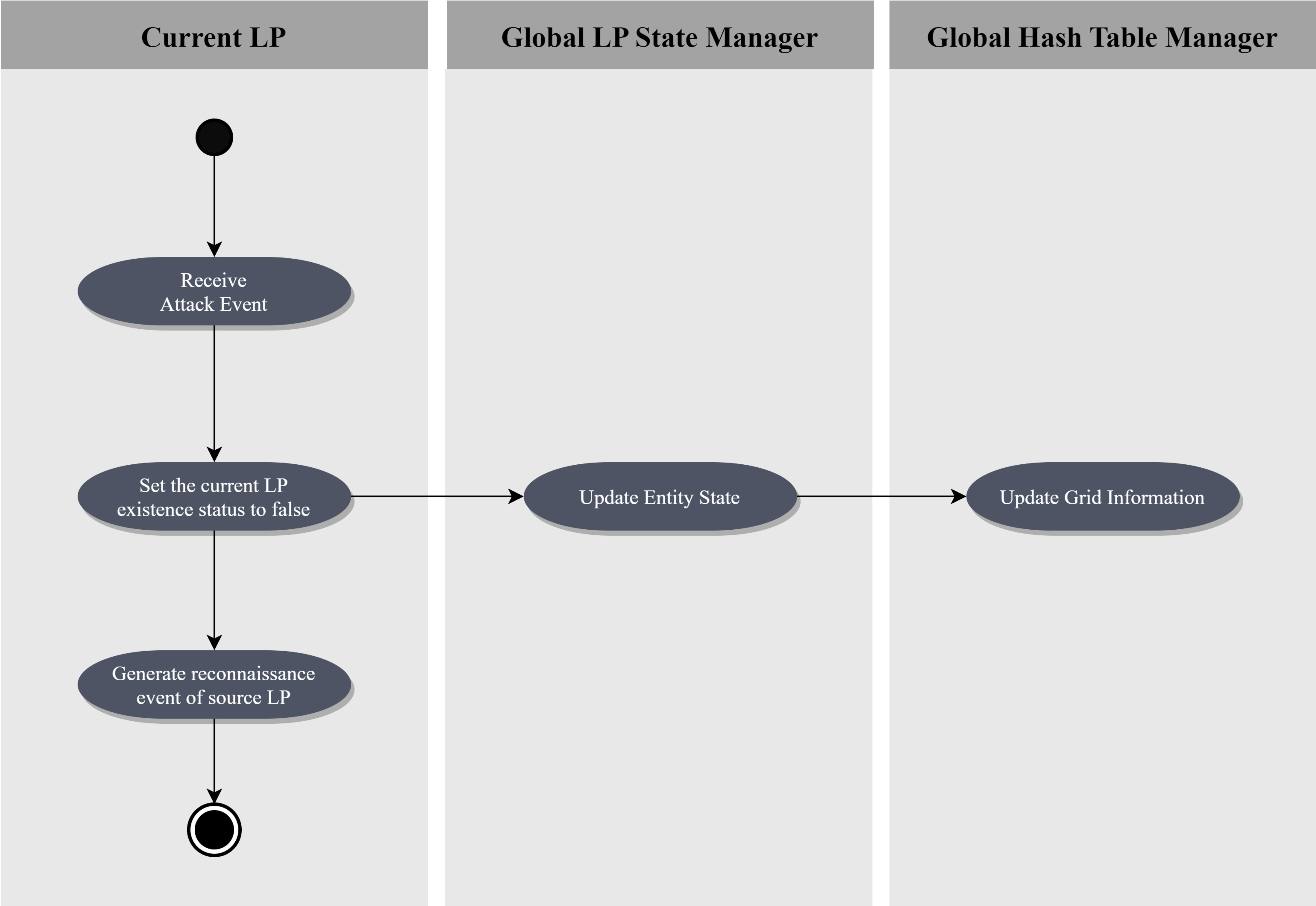}
        \caption{Logic of Attack Events}
        \label{tab:Logic of Attack Events}
    \end{figure}

    The attack event concludes the simulation loop of reconnaissance, missile movement, and attack. 
    It is triggered by a successful missile hit, setting the LP's existence status to false, indicating the entity's destruction, with the result logged in the data, as shown in Figure \ref{tab:Logic of Attack Events}.

\subsection{Large-scale Neighbor Entity Search Algorithm}
\label{section:Entity Search in Event Processing}
\begin{figure}[htbp]
        \centering
        \includegraphics[width=\textwidth]{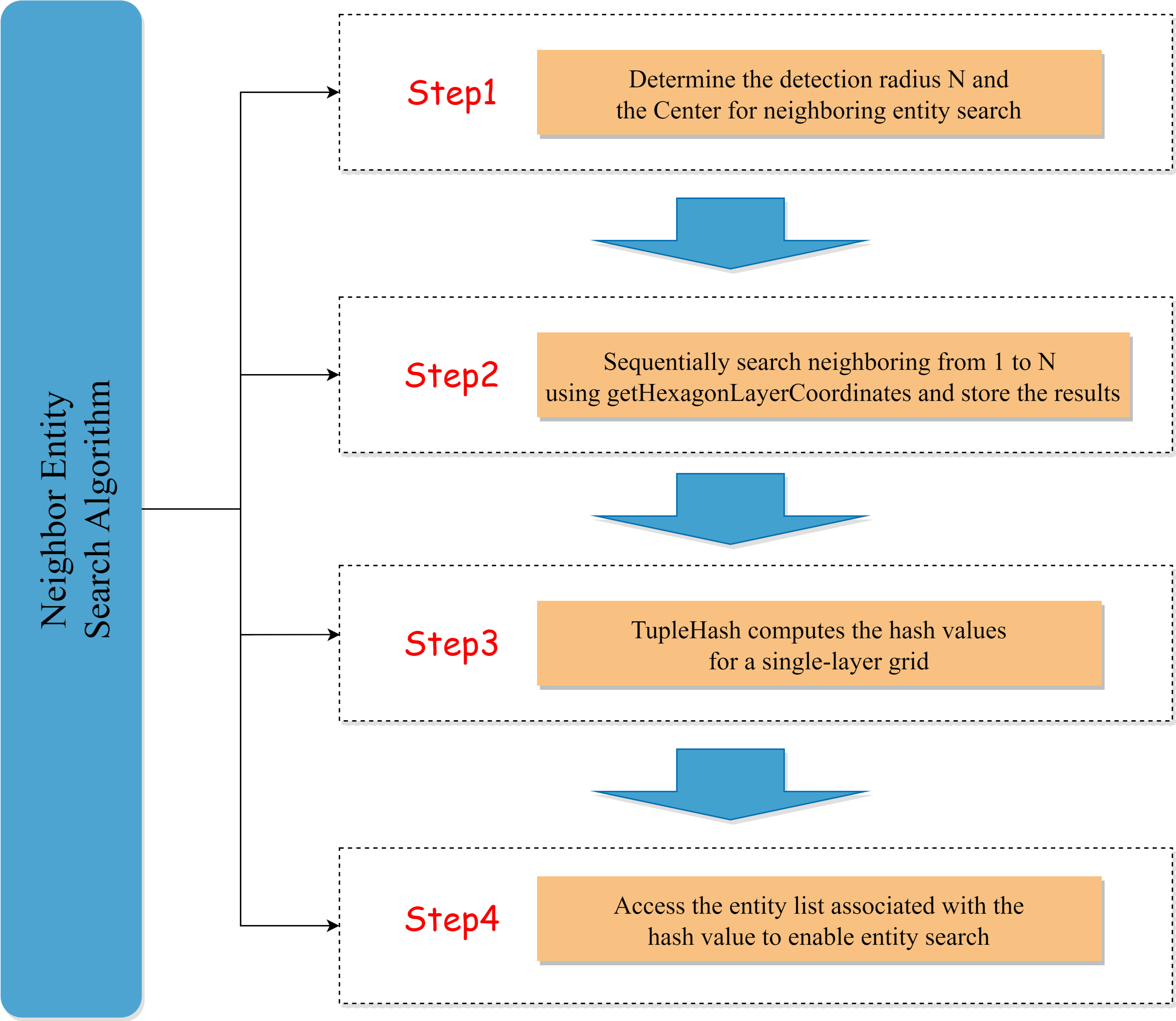}
        \caption{Steps for the Neighbor Entity Search Algorithm}
        \label{tab:Steps for the Neighbor Entity Search Algorithm}
    \end{figure}

    This section presents a single-level grid-based neighbor search algorithm, aimed at improving the efficiency of entity interaction, enhancing overall system performance, and reducing time costs. 
    The algorithm parallelizes the $gridMap\_$ and $gridToEntityMap\_$  hash tables in the GlobalDataSingleton and follows the steps outlined in Figure \ref{tab:Steps for the Neighbor Entity Search Algorithm} for neighbor search.
    
    We developed a custom hash function, TupleHash, to map grid data into a fixed-length hash value, which reduces the time complexity of neighbor searches to O($1$) and minimizes both data transmission and storage requirements. 
    As defined in Equation \ref{eq:HashValue}, $coord.x, coord.y$, and $coord.z$ represent the cube coordinates of the grid, while $p1, p2$, and $p3$ are large prime numbers. 
    HashSize is calculated as twice the hash table length plus one to ensure non-negativity of the result. 
    The computed hash value is stored in hashvalue, and linear probing is employed to resolve collisions, thereby improving efficiency and distribution. 
    The pseudocode for linear probing is provided in Algorithm \ref{alg:Linear probing method for handling hash collisions}.

    \begin{equation}
        \label{eq:HashValue}
        HashValue = (coord.x * P1 + coord.y * P2 + coord.z * P3 ) \% (HashValue) + 1
    \end{equation}

    Due to the issue of collisions in hash mapping, we employ linear probing to resolve hash conflicts. 
    Leveraging the average-case time complexity of O($1$) for neighborhood search algorithms and hash tables, we can reduce the time complexity of a single neighborhood entity search to O($1$). 
    Overall, the time complexity for the single-layer neighborhood search algorithm is O($n$), while the worst-case scenario for brute-force search is O($n^2$). 
    The neighborhood search algorithm only requires traversal of entities in the surrounding grid, enabling localized comparisons and enhancing search efficiency. 
    The neighborhood search entity algorithm is outlined in Algorithm \ref{alg:Search for neighboring entities}.

    \begin{algorithm}[htbp]
        \caption{Linear probing method for handling hash collisions}
        \label{alg:Linear probing method for handling hash collisions}
        {\bf Require:}\\ 
        \hspace*{0.05in} Input: $hashValue, Coordinate\& t$\\
        \hspace*{0.05in} Output: $hashValue$\\
        \begin{algorithmic}[1] 
            \State \textbf{Begin}
            \Function{findEmptySlot}{hashValue, t}
                \State index $\gets$ (hashValue \% TABLE\_SIZE) + 1
                \While{true}
                    \If{not table[index].occupied}
                        \State \Return index
                    \EndIf
                    \If{t == table[index].coordinate}
                        \State \Return index
                    \EndIf
                    \State index $\gets$ (index + 1) \% TABLE\_SIZE + 1
                    \If{index == (hashValue \% TABLE\_SIZE) + 1}
                        \State \Return TABLE\_SIZE
                    \EndIf
                \EndWhile
            \EndFunction
            \State \textbf{End}
        \end{algorithmic}
    \end{algorithm}

    \begin{algorithm}[H]
        \caption{Search for neighboring entities}
        \label{alg:Search for neighboring entities}
        {\bf Require:}\\ 
        \hspace*{0.05in} Input: $war\_event$\\
        \hspace*{0.05in} Output: $void$\\
        \begin{algorithmic}[1] 
            \Function{searchEntities}{war\_event}
            \State start $\gets$ getCurrentTime()
            \State center $\gets$ getEntityCenter()
            \State N $\gets$ getDetectionRadius()
            \For{i $\gets$ 1 \textbf{to} N}
                \State coordinates $\gets$ getHexagonLayerCoordinates(center, i)
                \For{coord \textbf{in} coordinates}
                    \State temp $\gets$ findIndexByCoordinate(coord)
                    \If{isValidIndex(temp)}
                        \For{entityPair \textbf{in} getEntitiesAtIndex(temp)}
                            \If{isValidTarget(entityPair)}
                                \State updateWarEvent(war\_event, entityPair)
                                \If{isMissile(entityPair)}
                                    \State \Return
                                \EndIf
                            \EndIf
                        \EndFor
                    \EndIf
                \EndFor
                \If{i == 1 \textbf{and} min\_distance\_name \textbf{is not} empty}
                    \State \Return
                \EndIf
            \EndFor
            \State end $\gets$ getCurrentTime()
            \State printDuration(start, end)
        \EndFunction
        \end{algorithmic}
    \end{algorithm}

\section{Experiments}

\subsection{Setup}

    Table \ref{tab:setup} documents the experimental setup configuration, detailing dual-socket AMD EPYC 7713 CPUs (128 cores), 1.5TB RAM, and verified software stack (Ubuntu 20.04.6/GCC 9.4.0/Python 3.8.10 with OpenMPI 4.0.3) to ensure computational reproducibility in distributed systems research.    

    \begin{table}[H]
        \centering
        \caption{System Configuration and Software Details}
        \label{tab:setup}
        \resizebox{\textwidth}{!}{  
        \begin{tabular}{ll}
        \toprule
        \textbf{Hardware Configuration} & \\
        \midrule
        \textbf{Server Model} & AS -4124GS-TNR \\
        \textbf{CPU} & 2 x AMD EPYC 7713 @ 1.70GHz (64 cores per CPU, 128 total cores) \\
        \textbf{Memory} & 1.5 TB \\
        \textbf{Storage} & 14.6 TB Disk \\
        \midrule
        \textbf{Software} & \\
        \midrule
        \textbf{Operating System} & Ubuntu 20.04.6 LTS \\
        \textbf{Containerization} & Docker 24.0.4 \\
        \textbf{Programming Environment} & Python 3.8.10, GCC 9.4.0 \\
        \textbf{Other Libraries} & OpenMPI 4.0.3, MPICH 4.1.2, glibc 2.31, \\
                                 & libcereal-dev 1.3.0-2, autoconf 2.71, \\
                                 & gcc-multilib 4:9.3.0-1, zlib 1:1.2.11.dfsg-2ubuntu1.5 \\
        \bottomrule
        \end{tabular}
        }
    \end{table}

\subsection{Effectiveness of load balancing based METIS}
    During the simulation, each unit processes events and generates new ones. 
    Initially, a static load-balancing strategy may seem adequate; 
    however, as the simulation runs, ensuring balanced computational loads across nodes becomes essential. 
    Previously, the event count and unit distribution were uniform enough that explicit load balancing was unnecessary. 
    However, practical scenarios introduce complexities: simulation units vary in characteristics and interaction frequencies. 
    Thus, dynamic load balancing is crucial for efficient performance.

    The previous implementation, using warped2's two-level data structure, avoids local event rollbacks by efficiently managing pending events. 
    This architecture also allows simple load balancing across threads: with multiple scheduling queues, each thread retrieves events from the queue mapped to its thread ID. 
    By dynamically remapping thread IDs to different queues on each retrieval, threads can draw from varied queues, distributing the workload more evenly.

    \begin{table}[H]
        \centering
        \caption{Simulation of Unit Migration Tests}
        \label{tab:unit_migration_tests}
        \begin{tabular}{cccc}
        \toprule
        Rank\_ID & Thread\_ID & Events & Events-Migration \\
        \midrule
        0 & 0 & 775000 & 525010 \\
        0 & 1 & 275020 & 525010 \\
        1 & 0 & 275000 & 275000 \\
        1 & 1 & 275000 & 275000 \\
        2 & 0 & 275000 & 275000 \\
        2 & 1 & 275000 & 275000 \\
        3 & 0 & 275000 & 275000 \\
        3 & 1 & 275000 & 275000 \\
        \midrule
        \multicolumn{2}{c}{Time} & 12.1238s & 10.6864s \\
        \bottomrule
        \end{tabular}
    \end{table}

    \begin{figure}[htbp]
        \centering
        \begin{minipage}[c]{0.45\textwidth}
            \centering
            \includegraphics[width=\textwidth]{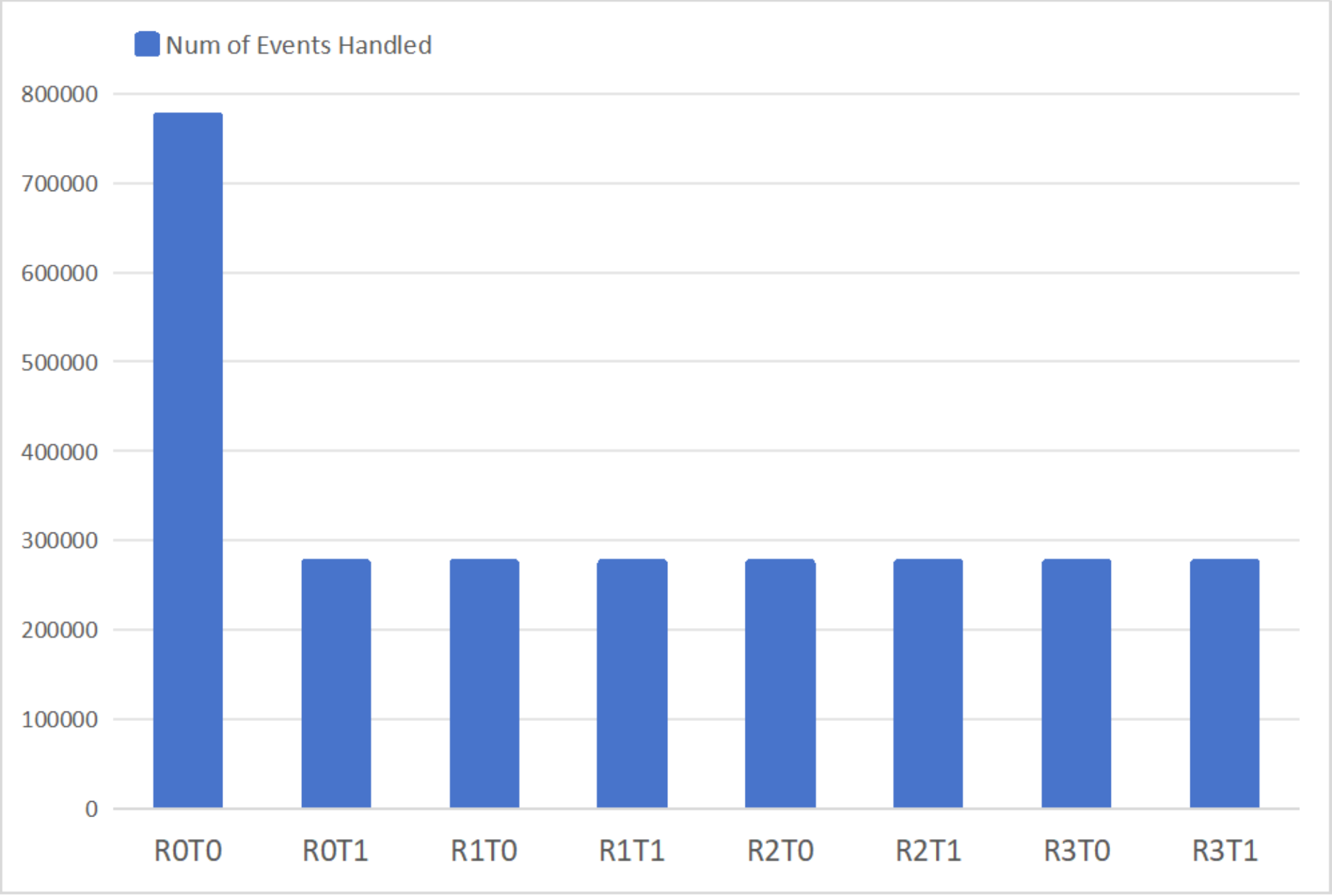}
            \subcaption{Analogue unit migration - OFF}
            \label{fig_17_1}
        \end{minipage} 
        \begin{minipage}[c]{0.45\textwidth}
            \centering
            \includegraphics[width=\textwidth]{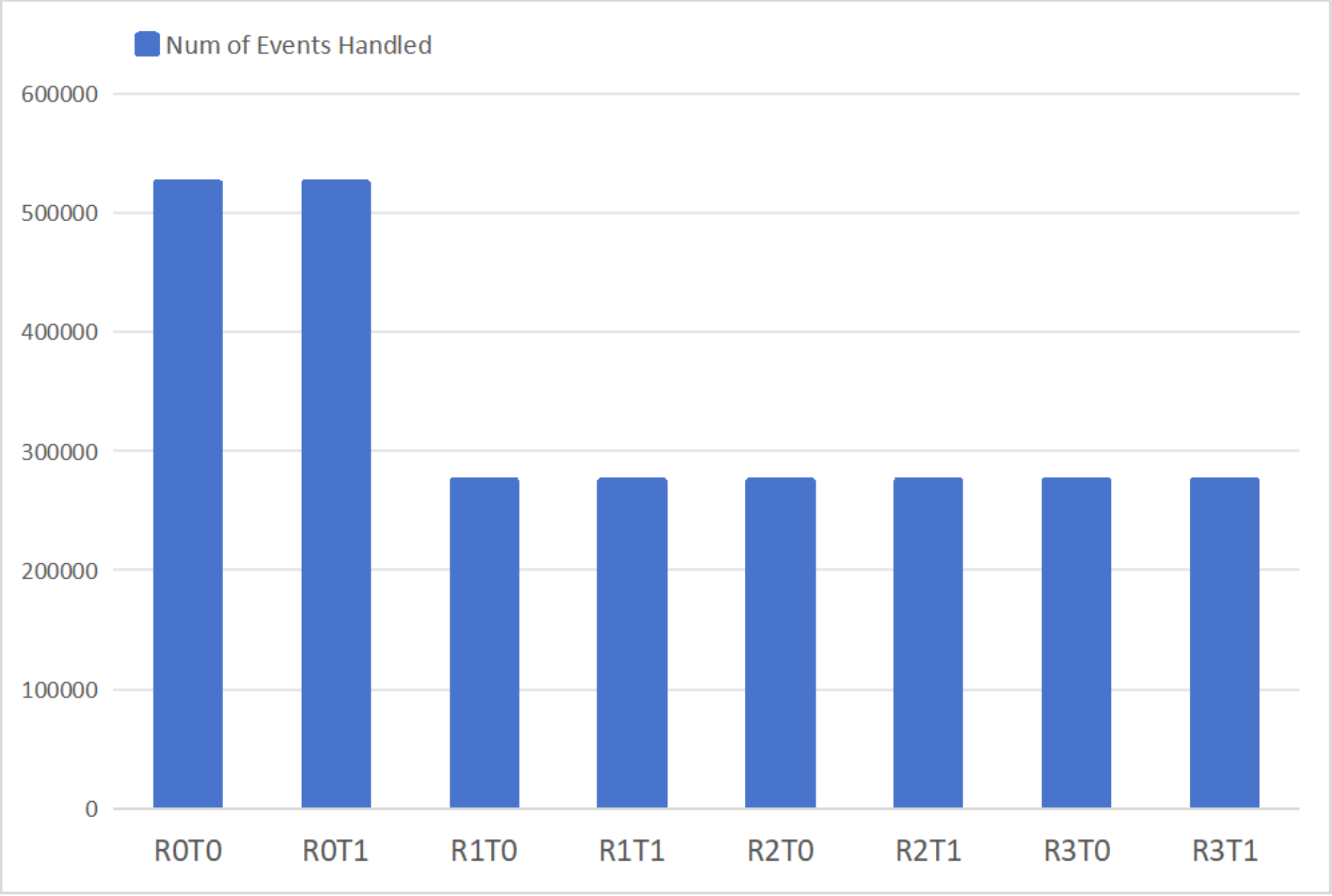}
            \subcaption{Analogue unit migration - ON}
            \label{fig_17_2}
        \end{minipage} 
        \caption{Unit Migration in Logical Process-Based Simulations}
        \label{fig_17}
    \end{figure}

    To validate the effectiveness of this mechanism, we constructed a test case involving 200,000 entities, each initialized with a single event, along with an additional 25,000 entities starting with three events. 
    Each event subsequently generates ten more events, simulating 10 time steps. 
    The simulation was executed on four MPI nodes, each configured with 2 worker threads. 
    The resulting data is shown in Table \ref{tab:unit_migration_tests} and Figure \ref{fig_17}.

    Balancing workloads solely within individual threads does not suffice; achieving optimal performance requires an even distribution of event processing across all threads globally. 
    While enabling migration of simulation units has yielded some performance improvement, significant imbalances remain, which continue to impact overall system efficiency.    

    \begin{table}[h!]
        \centering
        \caption{Load Balancing Optimisation Test}
        \label{tab:Load Balancing Optimisation Test}
        \begin{tabular}{ccccccc}
        \toprule
        Rank & Thread & Events & Migration & Balance & Balance-Migration \\ 
        \midrule
        0    & 0      & 274989 & 274994  & 204567    & 586417  \\
        0    & 1      & 275000 & 274995  & 968267    & 586417  \\
        1    & 0      & 275000 & 275000  & 678162    & 628062  \\
        1    & 1      & 275000 & 275000  & 577962    & 628062  \\
        2    & 0      & 275000 & 1525050 & 118393    & 570188  \\
        2    & 1      & 275100 & 1525050 & 1021982   & 570187  \\
        3    & 0      & 275000 & 275000  & 199078    & 565378  \\
        3    & 1      & 275000 & 275000  & 931678    & 565378  \\ 
        \midrule
        \multicolumn{2}{c}{Time}        & 34.938s & 26.527s & 15.016s  & 14.609s \\
        \multicolumn{2}{c}{SUM}         & \multicolumn{4}{c}{SAME} \\
        \bottomrule
        \end{tabular}
    \end{table}

    To address this, we begin by collecting initial load data from each simulation unit, establishing a baseline for global load distribution. 
    This data informs the partitioning of simulation units, aiming to achieve a balanced workload across the entire system. The warped2 framework's partitioning interface supports the necessary customization for this process. 
    Following a preliminary assessment, we adopted the METIS library integrated within warped2 to partition workloads for load balancing. Although METIS may not be the optimal tool, it was selected for its convenience, allowing us to verify the effectiveness of balanced partitioning prior to further optimization.
    To validate this approach, we constructed a scenario with pronounced load imbalance and measured execution times before and after partitioning. 
    The resulting data is illustrated in the accompanying Table \ref{tab:Load Balancing Optimisation Test}.

    \begin{figure}[H]
        \centering
        \includegraphics[width=\textwidth]{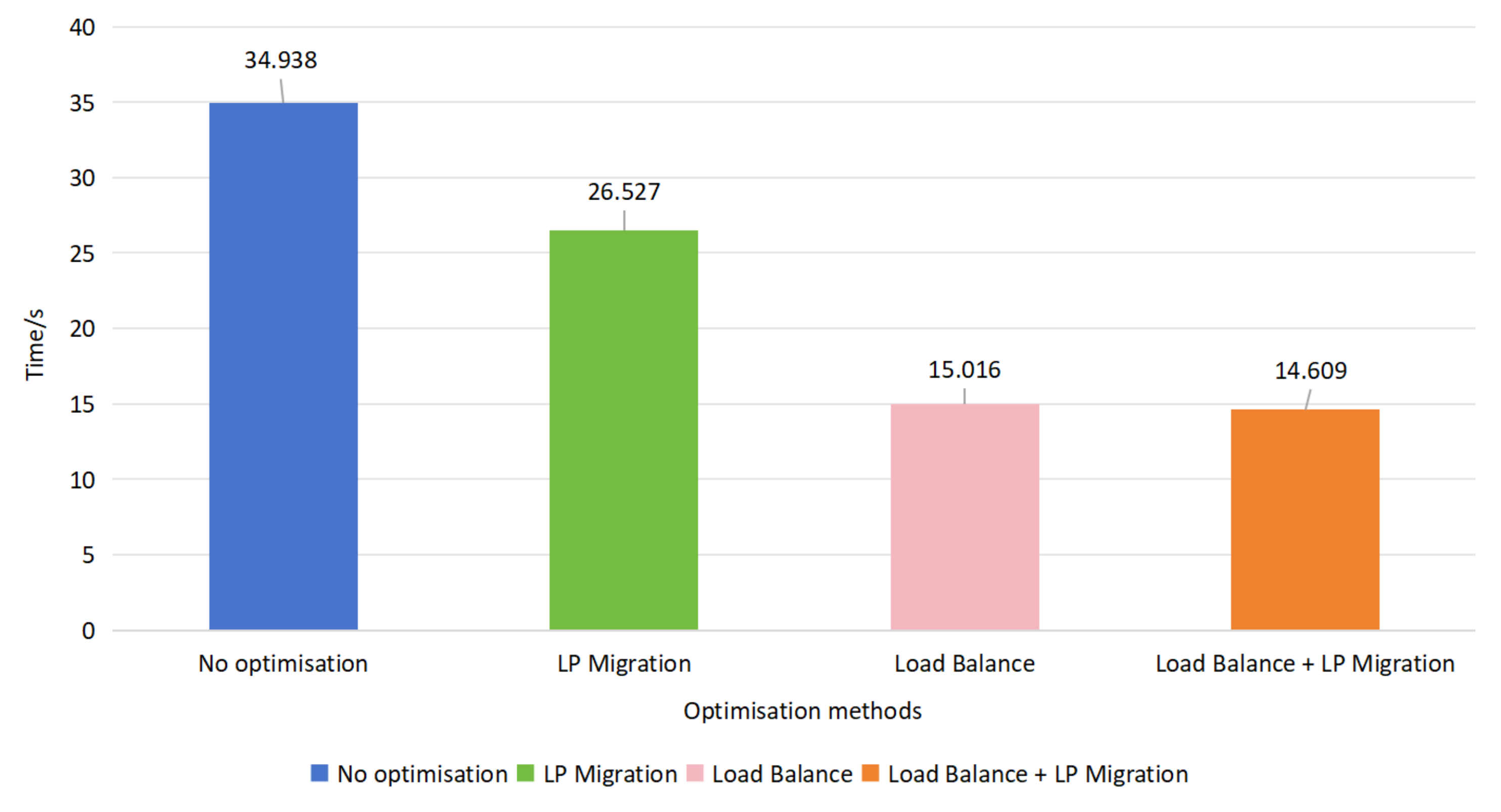}
        \caption{Dynamic Load Balancing Time Performance Comparison}
        \label{tab:Dynamic Load Balancing Time Performance Comparison}
    \end{figure}

    Figure \ref{tab:Dynamic Load Balancing Time Performance Comparison} and Table \ref{tab:Load Balancing Optimisation Test} indicate that applying two optimization techniques achieved the lowest execution time while balancing the load effectively across threads. 
    The combined optimization (Load Balance + LP Migration) achieves an execution time of 14.609 seconds, delivering an efficiency improvement of 58.18\% compared to the non-optimized baseline (34.938 seconds). Notably, Load Balance alone contributes 57\% of the total reduction, establishing it as the primary driver of the performance enhancement.
    The post-balancing performance improvements underscore the effectiveness of the METIS partitioning approach. 
    Partitioning was completed in under one second, well within acceptable limits.

    This analysis reveals that load balancing halved the execution time and balanced workloads across threads, aligning with parallel system expectations. 
    Future work will examine alternative parallel partitioning methods to further reduce partitioning overhead. 
    Additionally, we will explore the feasibility of dynamic load balancing to ensure sustained load balance and processing efficiency as the simulation progresses.

\subsection{Solver Correctness under Given rules}
    In this section, we conduct a rigorous evaluation of the correctness of the simulation approach based on derivation rules. The evaluation includes two main components: model logic and algorithm validation testing, and sensitivity analysis experiments, providing a comprehensive assessment of the simulation's accuracy.
    
    The model logic and algorithm validation tests confirm that the model operates as designed and that the search algorithms handle edge cases and invalid conditions correctly, ensuring consistent behavior and reducing failures. 
    We evaluated the model using four reliability metrics—success rate, average execution time, error rate, and relative standard deviation—across nine representative test cases. These metrics are calculated as follows: Success rate = (successful attempts / total cases) × 100\%; Average execution time = average time per test case; Error rate = (error occurrences / total cases) × 100\%; Relative standard deviation is computed using Equation \ref{eq:mean}, Equation \ref{eq:Standard Deviation}, and Equation \ref{eq:CV}.  
    \begin{equation}
        \label{eq:mean}
        \text{Mean} = \frac{\sum_{i=1}^{n} x_i}{n}
    \end{equation}
    
    \begin{equation}
        \label{eq:Standard Deviation}
        \text{Standard Deviation} = \sqrt{\frac{\sum_{i=1}^{n} (x_i - \text{Mean})^2}{n}}
    \end{equation}
    
    \begin{equation}
        \label{eq:CV}
        \text{CV} = \left( \frac{\text{Standard Deviation}}{\text{Mean}} \right) \times 100\%
    \end{equation}

    All simulations were conducted with a consistent entity scale of 100, ensuring that other objective factors remained unchanged. 
    The nine test cases were designed to encompass normal condition testing, critical condition testing, and abnormal invalid condition testing. 
    Each test case was repeated a minimum of five times, and the average values were computed. 
    The resulting data is presented in Table \ref{tab:Test Case Results}.

    \begin{table}[h!]
        \centering
        \caption{Test Case Results}
        \label{tab:Test Case Results}
        \resizebox{\textwidth}{!}{  
            \begin{tabular}{cccccc}
            \toprule
            Test Case & Success Rate (\%) & Average Search Time (ms) & Relative Standard Deviation (\%) & Error Rate (\%) \\
            \midrule
            Normal Search & 100 & 0.01650776 & 7.22449 & 0 \\
            Search Radius = 0 & 100 & 0.017685 & 6.08845 & 0 \\
            Search Radius = 1 & 100 & 0.008188 & 5.01719 & 0 \\
            Invalid Center Coordinate & 100 & 0 & 0 & 0 \\
            Negative Search Radius & 100 & 0 & 0 & 0 \\
            Empty war\_event & 100 & 0 & 0 & 0 \\
            Contains Invalid Temporary Entity & 100 & 0.0340425 & 9.58326 & 0 \\
            Process Hostile Units & 100 & 0.032 & 2.21224 & 0 \\
            Large Range Search & 100 & 0.0160142 & 1.73392 & 0 \\
            \bottomrule
            \end{tabular}
        }
    \end{table}

    Table \ref{tab:Test Case Results} shows the algorithm achieves a 100\% success rate with no errors. The average search time is under 0.035 ms, with fluctuations within 0.025 ms. Three test cases (e.g., empty war\_event) skipped the algorithm search step, confirming its accuracy in these scenarios.

    With a non-random seed, output consistency is less critical. Stability, assessed using the relative standard deviation (CV), remained below 10\% for all test cases, confirming the algorithm's robustness. These metrics validate the model's correctness and stability.

    Sensitivity analysis assessed how output changes with input variations. High sensitivity means small input changes lead to significant output fluctuations. We tested the number of persistent entities as input, and the scale and state of remaining entities as output. For each pair of adjacent entity counts, we recorded results under both parallel and sequential conditions.

    Each entity's state was represented by a vector of three attributes, normalized and simplified into a single value using a weighted sum.
    \begin{equation}
        \label{eq:State}
        \text{State}_i = [\text{Side}_i, \text{Type}_i, \text{Alive Status}_i]
    \end{equation}    
    \begin{equation}
        \label{eq:Normalized State_i}
        \text{Normalized State}_i = \frac{\text{State}_i - \text{Min}}{\text{Max} - \text{Min}}
    \end{equation} 
    \begin{equation}
        \label{eq:Y_State}
        Y_{\text{State}} = \frac{1}{N} \sum_{i=1}^{N} \text{Normalized State}_i
    \end{equation} 

    For each pair of adjacent entity counts ($X_1$ and $X_2$) and their corresponding remaining counts ($Y_{\text{count},1}$ and $Y_{\text{count},2}$) and states ($Y_{\text{state},1}$ and $Y_{\text{state},2}$), we calculate the variation using Equations \ref{eq:Delta X}, \ref{eq:Delta Y_count}, and \ref{eq:Delta Y_state}. 
    The results are then obtained by applying these values in Equations \ref{eq:S_count} and \ref{eq:S_state}. Table \ref{tab:Sensitivity Analysis Data} summarizes the calculated data.

    \begin{equation}
        \label{eq:Delta X}
        \Delta X = X_2 - X_1
    \end{equation}    

    \begin{equation}
        \label{eq:Delta Y_count}
        \Delta Y_{\text{count}} = Y_{\text{count},2} - Y_{\text{count},1}
    \end{equation} 

    \begin{equation}
        \label{eq:Delta Y_state}
        \Delta Y_{\text{state}} = Y_{\text{state},2} - Y_{\text{state},1}
    \end{equation} 

    \begin{equation}
        \label{eq:S_count}
        S_{\text{count}} = \frac{\Delta Y_{\text{count}} / Y_{\text{count},1}}{\Delta X / X_1}
    \end{equation} 

    \begin{equation}
        \label{eq:S_state}
        S_{\text{state}} = \frac{\Delta Y_{\text{state}} / Y_{\text{state},1}}{\Delta X / X_1}
    \end{equation} 
        
    \begin{table}[H]
        \centering
        \caption{Sensitivity Analysis Data}
        \label{tab:Sensitivity Analysis Data}
        \resizebox{\textwidth}{!}{
            \begin{tabular}{cccccccc}
            \toprule
            $X$ & Mode & $Y_{\text{count}}$ & $Y_{\text{state}}$ & $S_{\text{count}}$ (Lower Adjacent) & $S_{\text{count}}$ Relative & $S_{\text{state}}$ (Lower Adjacent) & $S_{\text{state}}$ Relative \\
            \midrule
            \multirow{2}{*}{500} & Serial & 26 & 0.4548 & 0.384615385 & \multirow{2}{*}{1.11\%} & 0.006299472 & \multirow{2}{*}{15.620\%} \\
              & Parallel & 27 & 0.4572 & 0.388888889 &   & 0.007283465 &   \\
            \multirow{2}{*}{1500} & Serial & 46 & 0.46053 & 0.619565217 & \multirow{2}{*}{5.92\%} & 0.015601589 & \multirow{2}{*}{25.605\%} \\
              & Parallel & 48 & 0.46386 & 0.65625 &   & 0.01959643 &   \\
            \multirow{2}{*}{2500} & Serial & 65 & 0.46532 & 0.507692308 & \multirow{2}{*}{5.62\%} & -0.00915499 & \multirow{2}{*}{11.108\%} \\
              & Parallel & 69 & 0.46992 & 0.536231884 &   & -0.010171944 &   \\
            \multirow{2}{*}{5000} & Serial & 98 & 0.46106 & 0.755102041 & \multirow{2}{*}{5.05\%} & -0.03574936 & \multirow{2}{*}{6.576\%} \\
              & Parallel & 106 & 0.46514 & 0.716981132 &   & -0.038100357 &   \\
            \multirow{2}{*}{7500} & Serial & 135 & 0.4528187 & 0.866666667 & \multirow{2}{*}{5.77\%} & -0.00264808 & \multirow{2}{*}{10.615\%} \\
              & Parallel & 144 & 0.456279 & 0.916666667 &   & -0.002366973 &   \\
            \multirow{2}{*}{10000} & Serial & 174 & 0.452419 & 0.762452107 & \multirow{2}{*}{4.19\%} & 0.008282735 & \multirow{2}{*}{8.394\%} \\
              & Parallel & 188 & 0.455919 & 0.730496454 &   & 0.007587459 &   \\
            \multirow{2}{*}{25000} & Serial & 373 & 0.4580399 & 0.694369973 & \multirow{2}{*}{3.14\%} & -0.002257445 & \multirow{2}{*}{11.824\%} \\
              & Parallel & 394 & 0.4611079 & 0.672588832 &   & -0.002524355 &   \\
            \multirow{2}{*}{50000} & Serial & 632 & 0.4570059 & - & - & - & - \\
              & Parallel & 659 & 0.4599439 & - & - & - & - \\
            \bottomrule
            \end{tabular}
        }
    \end{table}
    
    We find that $S_{\text{count}}$ ranges from 0 to 1, indicating stability in the simulation model at the entity scale, which aids in effective battlefield analysis while minimizing unnecessary assessments. 
    The relative error of $S_{\text{count}}$ is between 1\% and 6\%, mostly clustering around 3\% to 5\%. At 500 entities, the relative error is 1.11\%, showing low error. 
    While the error slightly increases with more entities, it remains low overall, confirming the accuracy and consistency of both parallel and serial simulation outputs.
    The experimental data addresses the following issues:
    \begin{enumerate}[(1)]       
        \item Trend of Error Fluctuations: As the number of permanent entities increases, relative errors in $S_{\text{count}}$ and $S_{\text{state}}$ fluctuate more at smaller scales but stabilize at larger scales, indicating greater sensitivity at smaller scales and improved stability with more entities.
        \item Impact of System Scale and Complexity on State and Error: Table \ref{tab:Sensitivity Analysis Data} shows that $S_{\text{state}}$ generally has higher errors than $S_{\text{count}}$ at smaller scales due to randomness. Larger scales reduce the error in $S_{\text{state}}$, indicating better stability and lower complexity, with smaller scales showing higher sensitivity and operational issues.
    \end{enumerate}

\subsection{System Performance Under Different Search Strategies}
\begin{figure}[H]
        \centering
        \includegraphics[width=\textwidth]{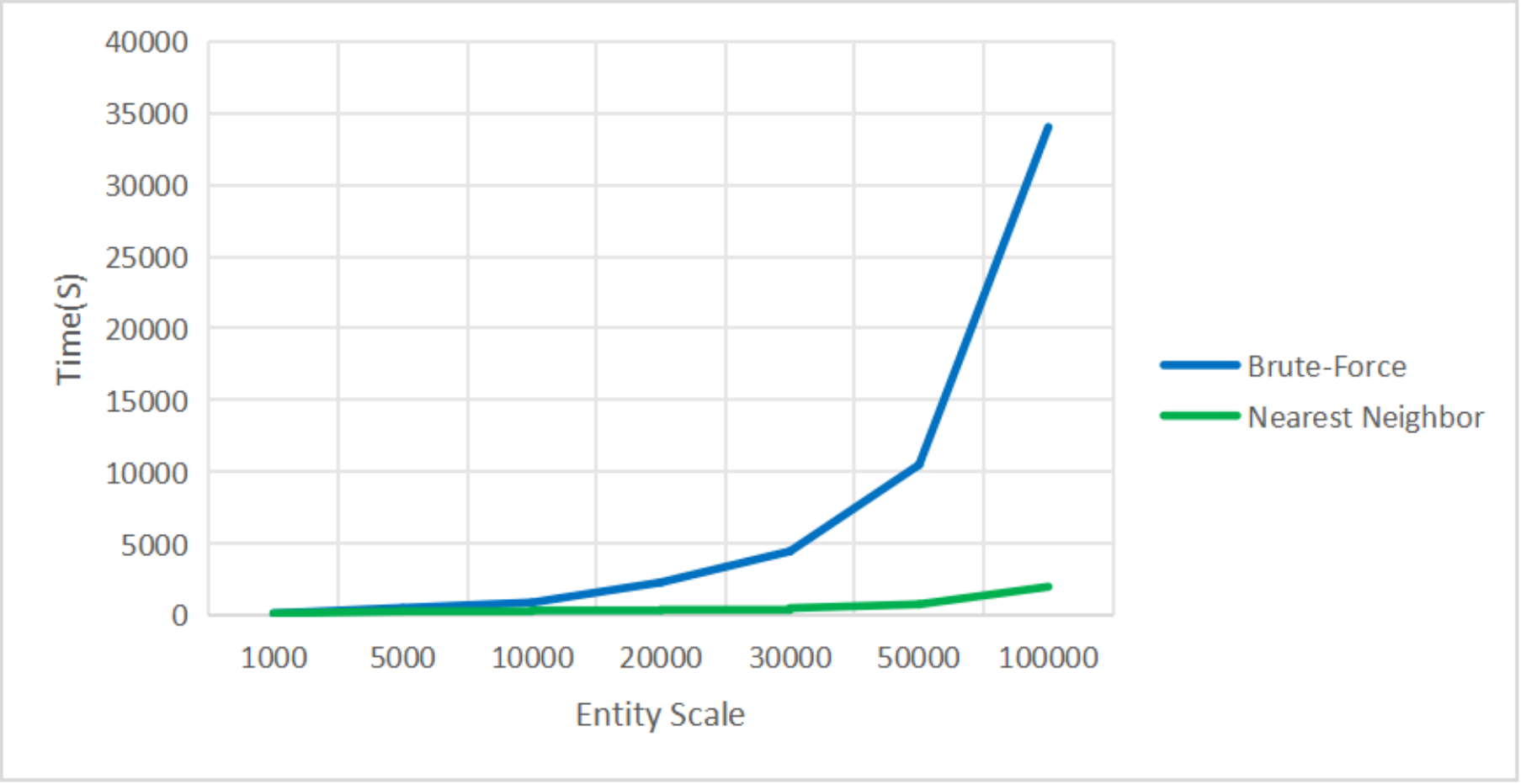}
        \caption{Times for MPI parallelization}
        \label{fig:Times for MPI parallelization}
    \end{figure}

\begin{figure}[H]
        \centering
        \includegraphics[width=\textwidth]{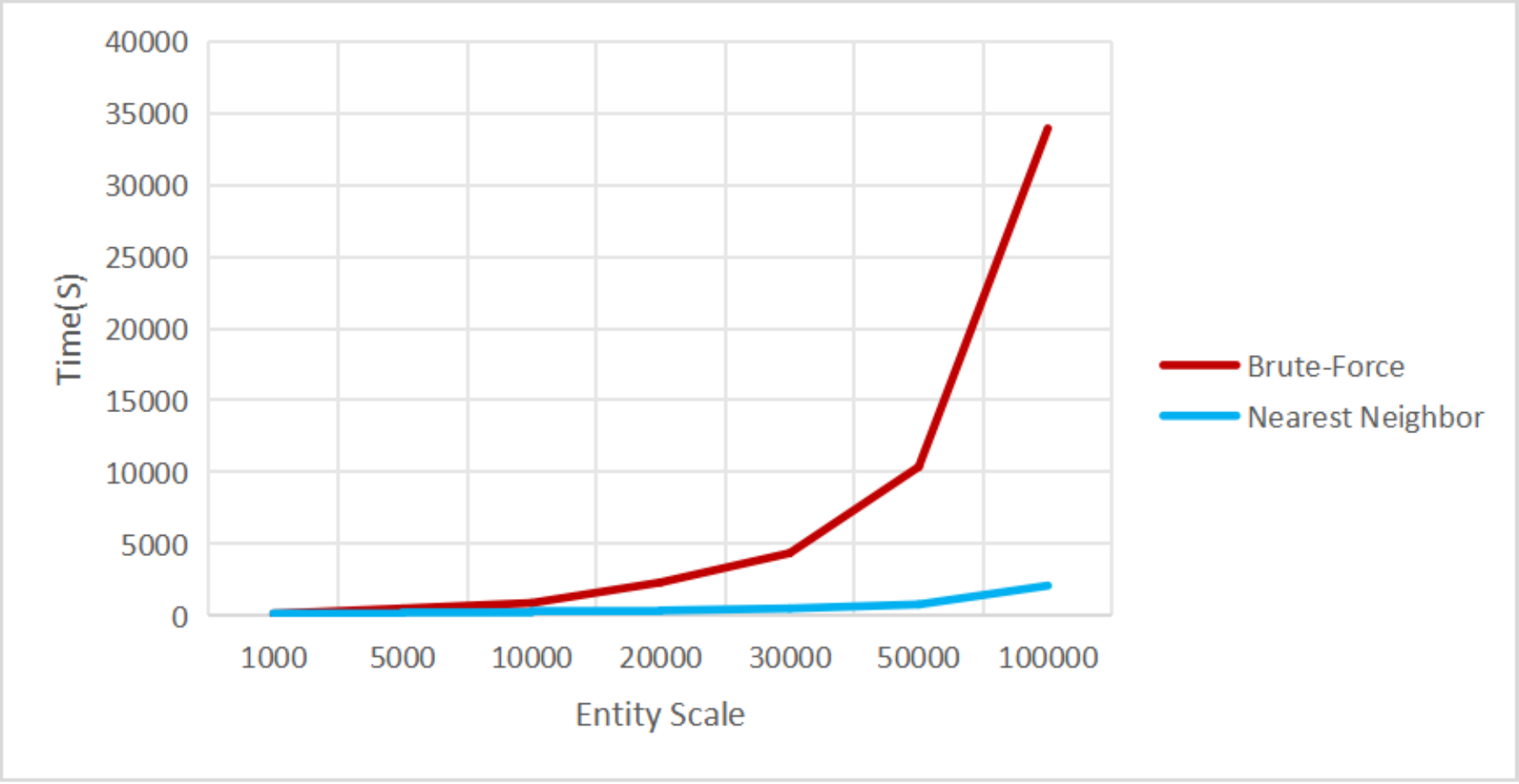}
        \caption{Times for Pthreads parallelization}
        \label{fig:Times for Pthreads parallelization}
    \end{figure}

    In parallel computing, two common parallelization approaches are employed: inter-process shared memory parallelization based on MPI and intra-thread shared memory parallelism based on Pthreads. 
    This section evaluates the simulation times of different search algorithms across varying entity scales for both parallel strategies, with a maximum simulation duration set at 10,000 ms. 
    By comparing the actual simulation times in different scenarios, we assess the performance of various algorithms. 
    The detailed data is presented in Figures \ref{fig:Times for MPI parallelization} and \ref{fig:Times for Pthreads parallelization}.

    In Figure \ref{fig:Times for MPI parallelization}, under the MPI single-core dual-thread configuration, data exchange between threads is facilitated. 
    At lower entity scales, the time consumption of the Brute-Force Search and Nearest Neighbor Search algorithms is comparable. 
    Due to the time complexity of the Brute-Force Search being $O(n^2)$ and that of the Nearest Neighbor Search being $O(n)$, the time consumption of the Brute-Force Search increases exponentially as the entity scale rises. 
    This is clearly illustrated in Figure \ref{fig:Times for MPI parallelization}: starting from an entity scale of 10,000, the simulation speed for the Nearest Neighbor case is approximately three times that of the Brute-Force case. 
    For entity scales of 20,000 to 30,000, the speed improvements for the Nearest Neighbor scenario relative to the Brute-Force scenario are 576\% and 861.2\%, respectively. 
    For scales of 50,000 to 100,000, the improvements are 1336.5\% and 1694.1\%, respectively, with enhancements becoming more pronounced at larger entity scales.
    
    In Figure \ref{fig:Times for Pthreads parallelization}, the simulator employs Pthreads in a single-core dual-thread configuration, yielding speed improvements comparable to those achieved with MPI. 
    For entity scales of 20,000 to 30,000, the speed enhancements for the Nearest Neighbor case relative to the Brute-Force Search are 629.4\% and 855.28\%, respectively. 
    For scales of 50,000 to 100,000, the improvements are 1315.1\% and 1618.75\%, respectively.

    Based on the previous tests, we observed that the simulation times for the two parallel strategies are largely consistent. 
    However, we lack metrics to quantitatively demonstrate the performance improvement of parallel execution relative to serial execution and the enhancement provided by different search strategies. 
    Thus, we introduce two new metrics: speedup and efficiency. 
    We computed the speedup for various scenarios across different entity scales, as detailed in Equations \ref{eq:S_Brute Force}, \ref{eq:S_Neighbor}, and \ref{eq:S_Relative}.

    \begin{equation}
        \label{eq:S_Brute Force}
        S_{\text{Brute Force}} = \frac{\text{Brute Force Serial Simulation Time}}{\text{Brute Force Parallel Simulation Time}}
    \end{equation} 

    \begin{equation}
        \label{eq:S_Neighbor}
        S_{\text{Neighbor}} = \frac{\text{Neighbor Search Serial Simulation Time}}{\text{Neighbor Search Parallel Simulation Time}} 
    \end{equation} 

    \begin{equation}
        \label{eq:S_Relative}
        S_{\text{Relative}} = \frac{\text{Brute Force Parallel Simulation Time}}{\text{Neighbor Search Parallel Simulation Time}}
    \end{equation} 

\begin{figure}[htbp]
        \centering
        \begin{minipage}[c]{0.75\textwidth}
            \centering
            \includegraphics[width=\textwidth]{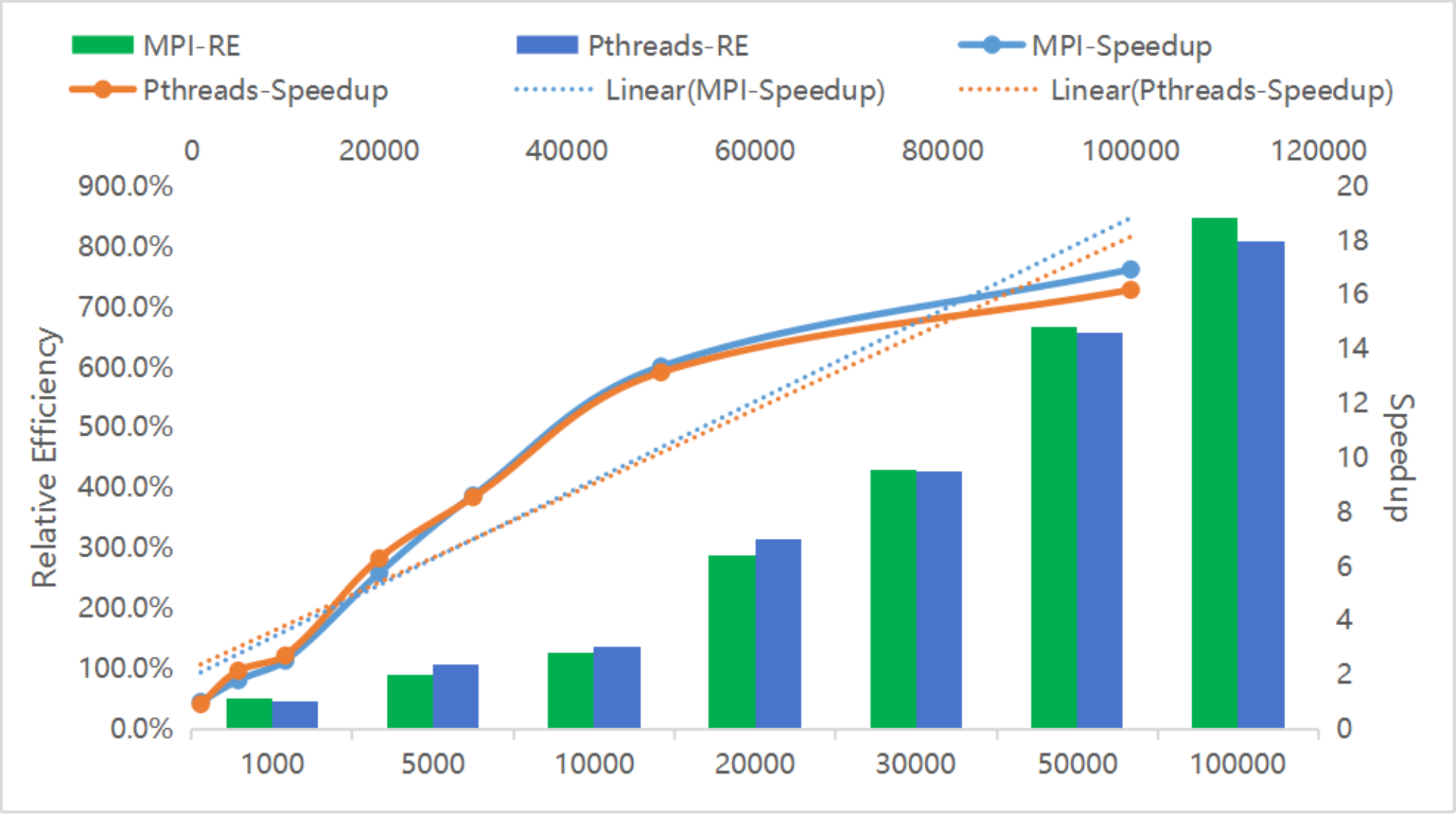}
            \subcaption{Relative speedup and efficiency of different parallel strategies under various algorithms}
            \label{fig:E2_1}
        \end{minipage} \\
        \begin{minipage}[c]{0.75\textwidth}
            \centering
            \includegraphics[width=\textwidth]{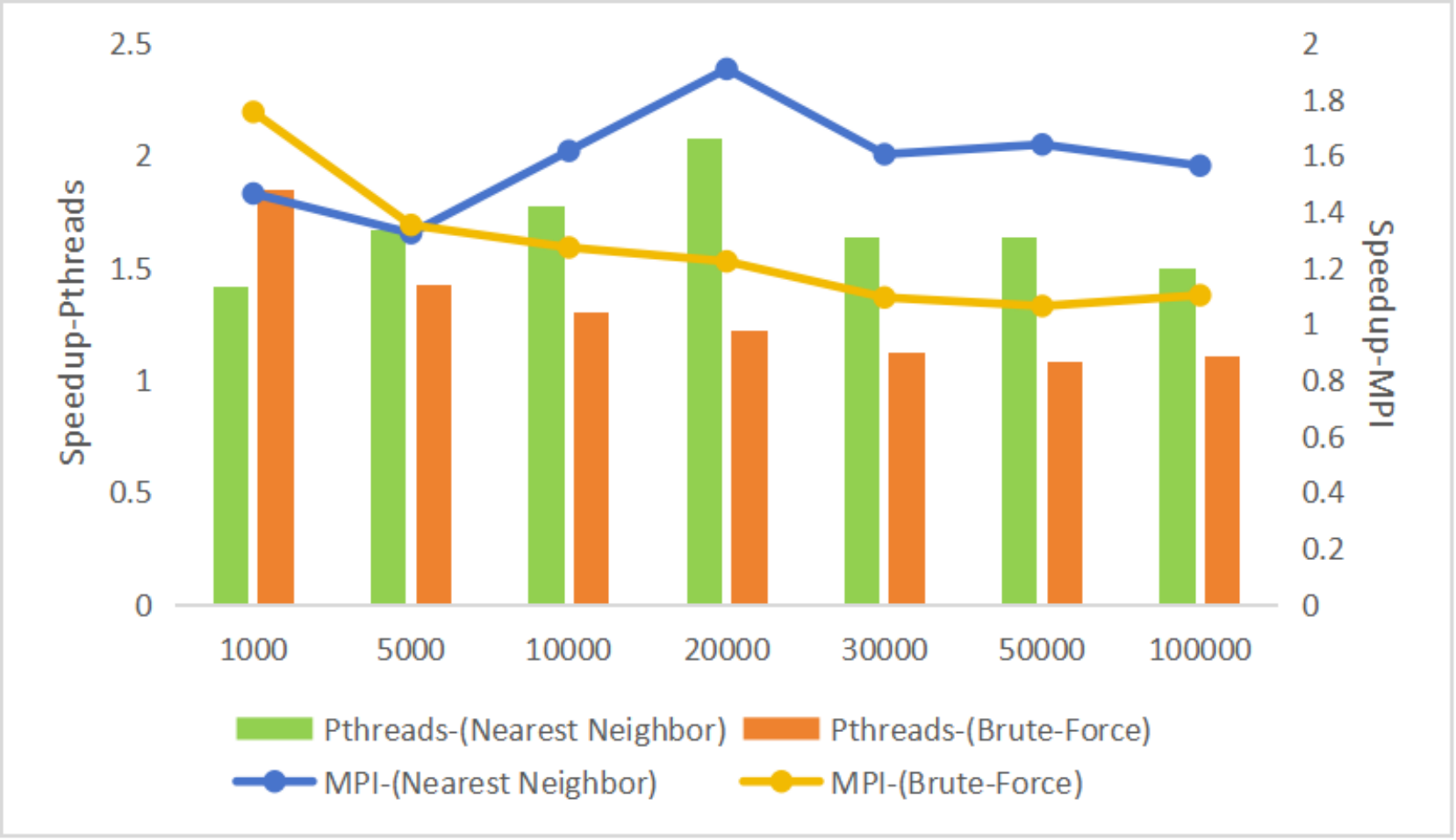}
            \subcaption{Comparison of speedup ratios across different algorithms and parallel strategies based serial}
            \label{fig:E2_2}
        \end{minipage} 
        \begin{minipage}[c]{0.75\textwidth}
            \centering
            \includegraphics[width=\textwidth]{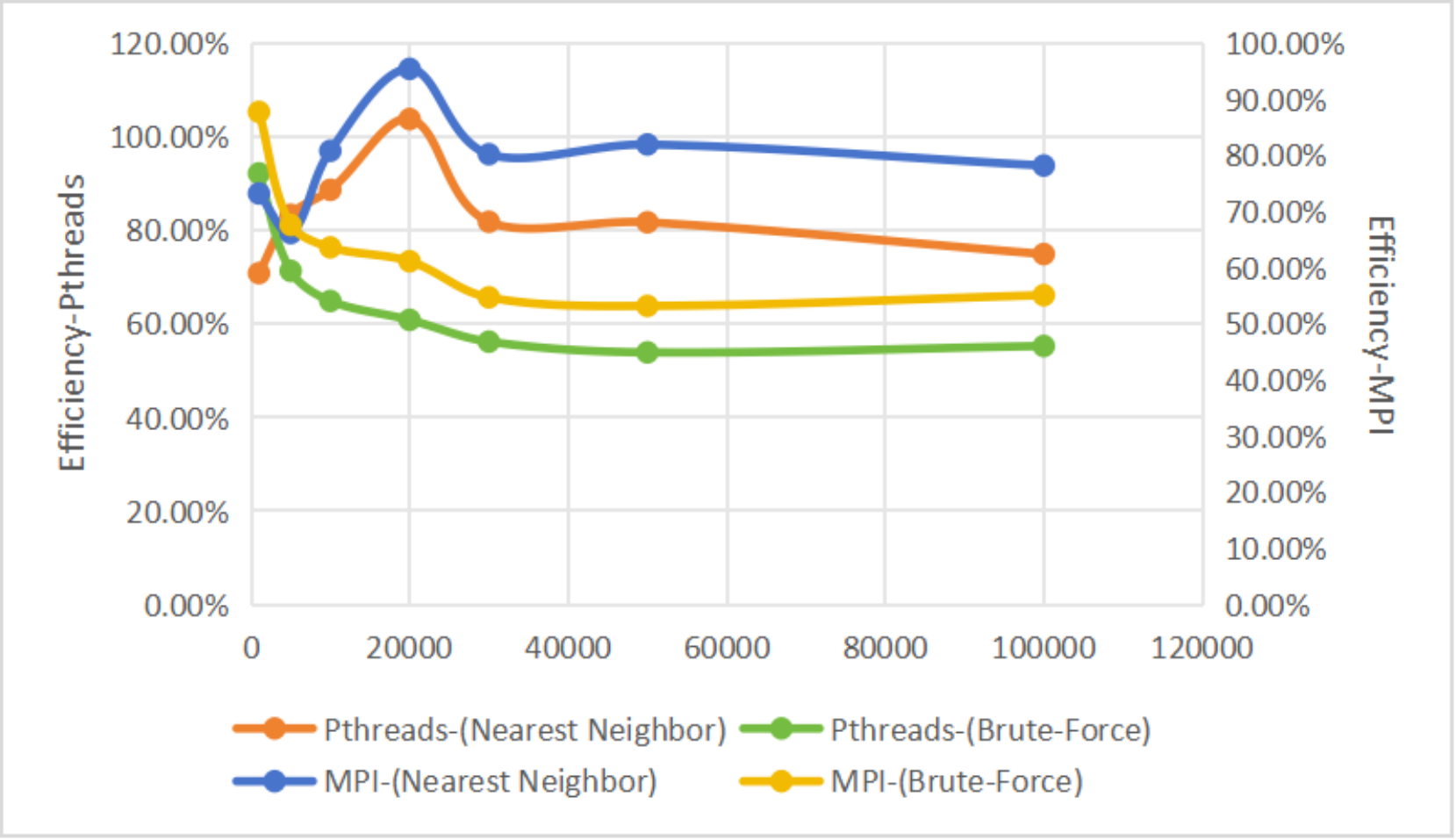}
            \subcaption{Comparison of efficiency ratios across different algorithms and parallel strategies based serial}
            \label{fig:E2_3}
        \end{minipage}
        \caption{Speedup and Efficiency under different conditions}
        \label{fig:Speedup and Efficiency under different conditions}
    \end{figure}

    We computed the speedup for three scenarios, comparing the performance improvements of parallel execution versus serial execution and neighbor search relative to brute force search. 
    For efficiency calculations, we used relative efficiency for the performance enhancement of neighbor search compared to brute force, while other efficiency metrics are provided in Equation \ref{eq:Efficiency}. 
    With a fixed parallelism degree of $P = 2$, the results are presented in Figure \ref{fig:Speedup and Efficiency under different conditions}.

    \begin{equation}
        \label{eq:Efficiency}
        \text{Efficiency} = \frac{S}{\text{Parallelism Degree (P)}}
    \end{equation} 

    From Figure \ref{fig:E2_1}, it is evident that the speedup achieved with MPI increases significantly from 1.006 to 16.941 as the problem size grows, indicating that the nearest neighbor search strategy in the MPI context can effectively leverage parallel processing capabilities. 
    Similarly, the speedup for Pthreads follows a comparable trend, although its growth rate is slightly lower than that of MPI, particularly at smaller scales, where the speedup for Pthreads is below that of MPI. 
    The relative efficiency of MPI starts at 50.3\% and gradually increases to 847.1\%, demonstrating superior capability in handling large-scale problems, especially when data sizes reach 30,000 or more, where a relative efficiency close to 430.6\% indicates effective resource utilization. 
    The relative efficiency for Pthreads also exhibits a similar trend, beginning at 46.3\% and peaking at 809.4\%. Although its relative efficiency is slightly lower than that of MPI, it still shows commendable performance at larger scales. 
    This clearly illustrates that both parallel strategies in the context of nearest neighbor search demonstrate a superior ability to utilize parallel resources compared to brute-force search, thereby enhancing system simulation performance.

    Figures \ref{fig:E2_2} and \ref{fig:E2_3} show the performance gains of parallel over serial processing for different search strategies and parallel methods, measured by speedup and efficiency. 
    Across all parallel strategies, parallel processing substantially outperforms serial processing. 
    Nearest Neighbor search consistently achieves greater speedup and efficiency compared to brute-force search. 
    In large-scale MPI parallelization, neighbor search outpaces brute-force by an average speedup of 0.6 and 30\% increase in efficiency, with similar trends observed in Pthreads parallelization, as confirmed in Figure \ref{fig:E2_3}.

    \begin{figure}[H]
        \centering
        \includegraphics[width=\textwidth]{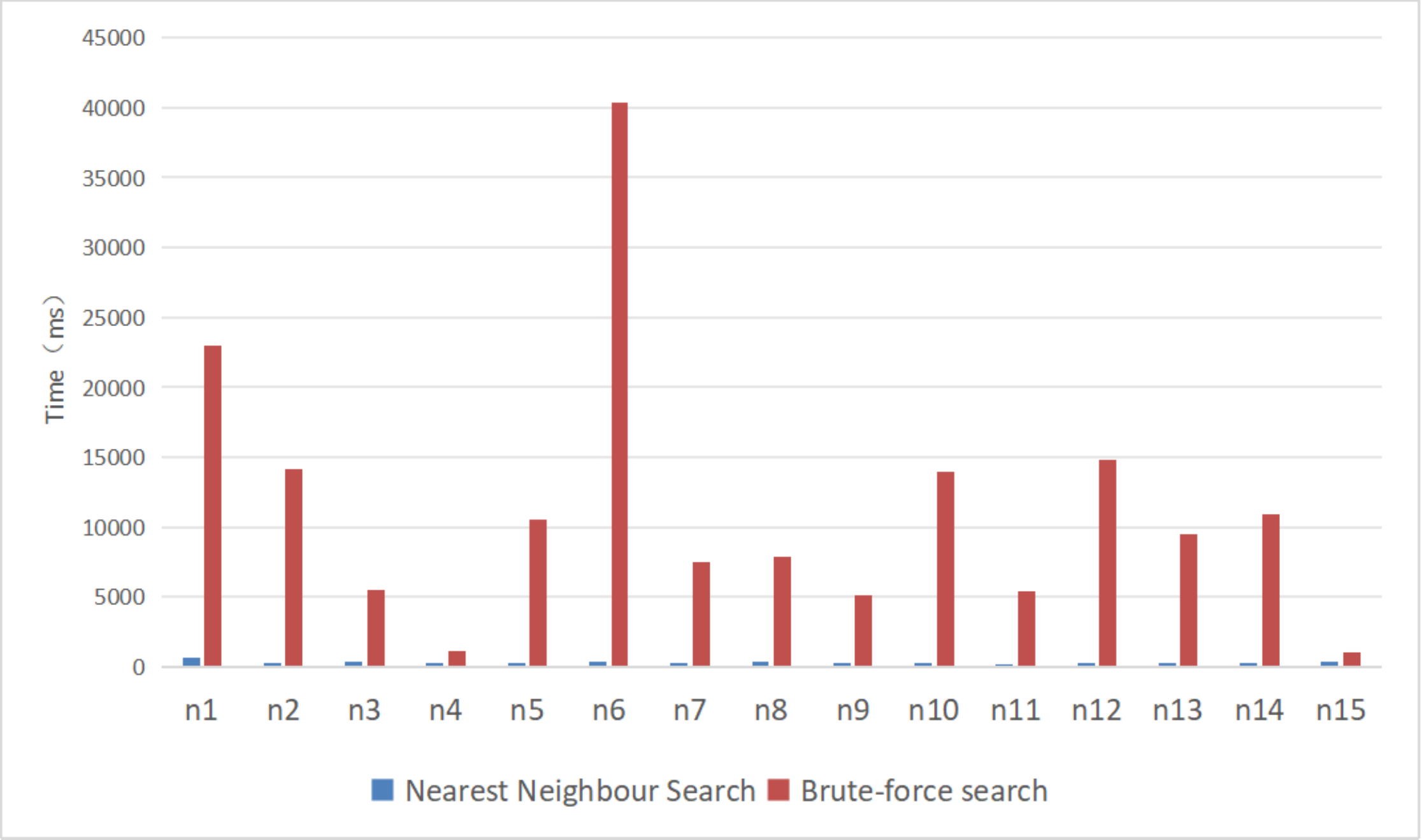}
        \caption{Comparison of single run for different searches at entity scale 100,000}
        \label{fig:Comparison of single run for different searches at entity scale 100,000}
    \end{figure}

    The impact of entity scales and parallel strategies on simulation performance is significant. 
    Additionally, we compared the performance of different nearest neighbor search algorithms in Figure \ref{fig:Comparison of single run for different searches at entity scale 100,000}. 
    At smaller scales, the difference is minimal, but at 100,000 entities, the nearest neighbor search outperforms brute-force search. 
    The nearest neighbor search had an average runtime of 326.53 microseconds, compared to 11,382.33 microseconds for brute-force, showing a 3,485.81\% performance improvement and demonstrating its efficiency in large-scale simulations.

\subsection{System Validation: GridWorld Demo}
    Through the previous experimental validation, we have confirmed the performance improvements and correctness of the solver under the optimized framework, with clear evidence of increased processing efficiency. 
    However, to assess the operability of the large-scale parallel discrete event simulation engine based on Warped2, a more concrete system validation of the simulation process and result data is required.
    
    \begin{figure}[H]
        \centering
        \includegraphics[width=0.75\textwidth]{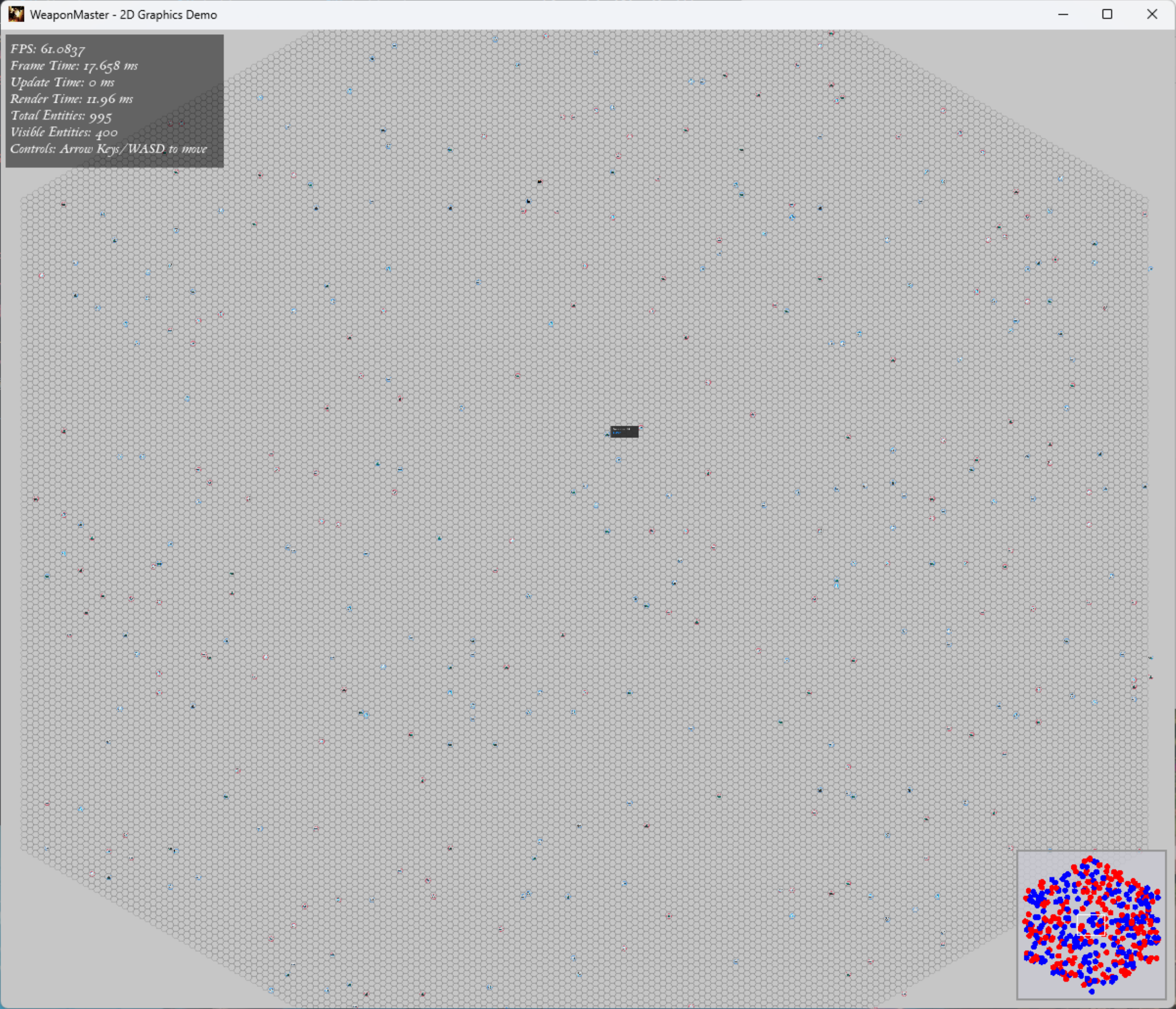}
        \caption{Framework Performance Analysis: 2D Rendering with Interactive Entity Visualization}
        \label{fig:UI}
    \end{figure}

    To achieve this, we developed a Gridworld Demo using the SFML visualization library for system validation as shown in Figure \ref{fig:UI}. By implementing a custom event output structure EntityInfo, within the simulator, we can extract event data. 
    As the events represent LP state changes, storing event information at each timestamp enables us to visually reconstruct the simulation's interactions in the SFML frontend.
    The interface adopts a gray hexagonal honeycomb grid as the foundational background layout, with entities from distinct factions annotated using high-contrast color coding (blue and red) to enhance visual discrimination. 
    The interface is partitioned into three functional quadrants dedicated to real-time status identification, performance metric monitoring, and tactical situation visualization respectively.
    \begin{figure}[htbp]
        \centering
        \includegraphics[width=\textwidth]{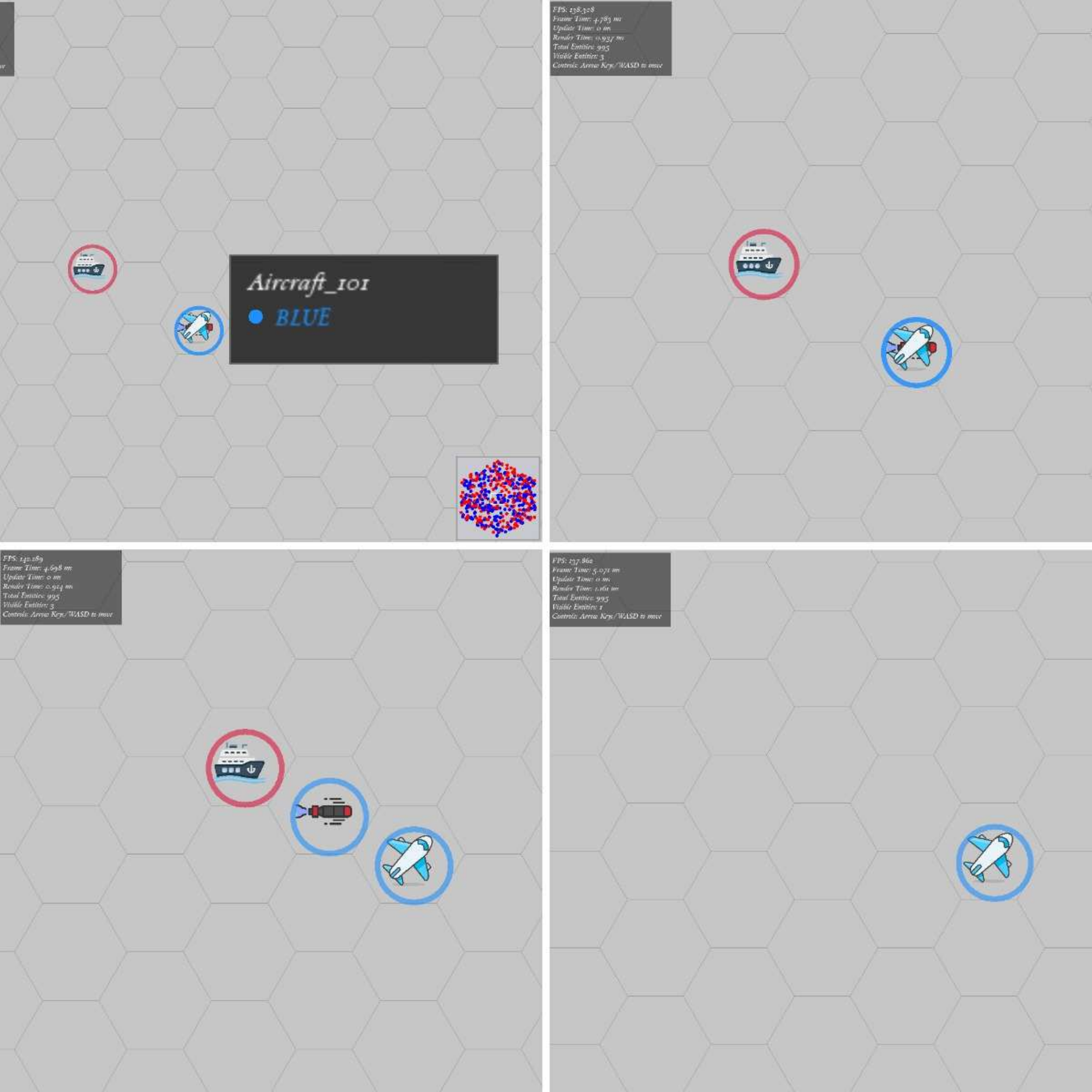}
        \caption{Representation of the entity interaction process}
        \label{fig:Representation of the entity interaction process}
    \end{figure}

    Figure \ref{fig:Representation of the entity interaction process} shows a partial visualization of the entity interaction cycle, illustrating the aircraft's reconnaissance of enemy tanks, missile move, attack and the destruction of the enemy entities.
    
\begin{figure}[htbp]
        \centering
        \includegraphics[width=\textwidth]{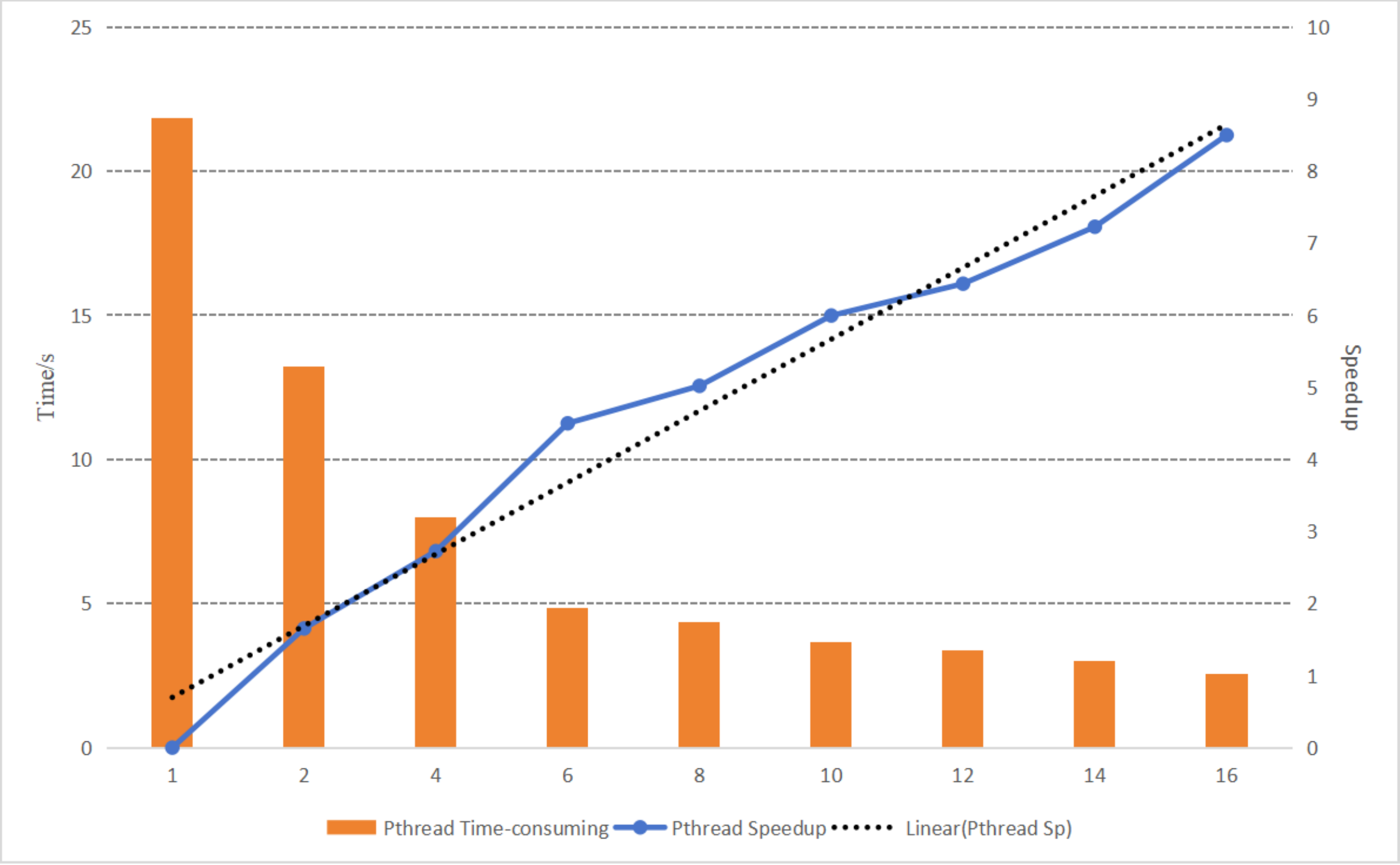}
        \caption{Pthreads Time and Speedup vs Processes}
        \label{fig:Pthreads Time and Speedup vs Processes}
    \end{figure}

\begin{figure}[htbp]
        \centering
        \includegraphics[width=\textwidth]{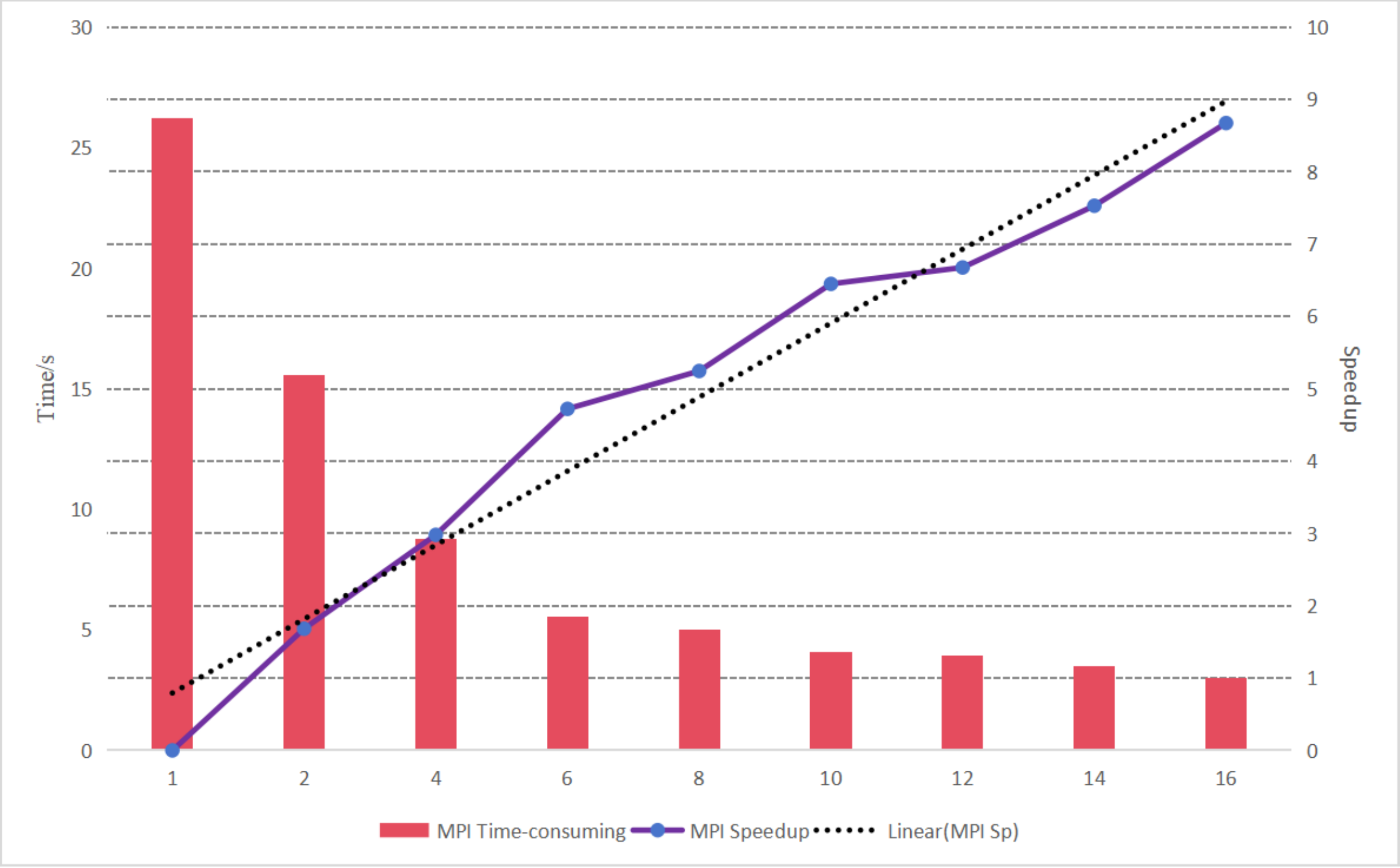}
        \caption{MPI Time and Speedup vs Processes}
        \label{fig:MPI Time and Speedup vs Processes}
    \end{figure}

    Figure \ref{fig:Pthreads Time and Speedup vs Processes} and \ref{fig:MPI Time and Speedup vs Processes} compares the speedup ratios and computational time consumption (in seconds) of Pthread and MPI under parallel granularities ranging from 1 to 16 within our implemented engine framework GridWorld Demo for large-scale distributed parallel discrete event simulation, 
    where performance testing was conducted under a GVT of 10, testing with an entity scale of 20k and identical nearest-neighbor search algorithm configurations. 

    The Pthread strategy demonstrates superior speedup performance, scaling from a baseline speedup of 1 (single-thread execution time: 21.85s) to 8.50× (2.57s at 16 threads), achieving 53.1\% of the theoretical linear speedup (ideal 16× for 16 threads), while outperforming MPI in low-to-medium concurrency ranges (1–8 threads) with an 8-thread speedup of 5.02× compared to MPI's 5.25×, accompanied by an 88.2\% reduction in execution time from 21.85 seconds to 2.57 seconds. 
    
    In contrast, MPI exhibits enhanced scalability in high-concurrency regimes (12–16 tasks), attaining a speedup of 8.67× (54.2\% of theoretical value) at 16 tasks, marginally surpassing Pthread's 8.50×, with execution time decreasing from 26.21 seconds (single-process) to 3.02 seconds (88.5\% reduction), reflecting slower scalability degradation in distributed environments due to reduced resource contention.

    Based on the previously discuss high throughput performance, simulation framework correctness, and performance test results, the system has been successfully validated. 
    The optimization of kernel load balancing and high throughput has been achieved, and the solver has been developed and applied while ensuring correctness. 
    The simulation system's performance has been significantly improved, demonstrating that the proposed simulation framework can be extended to real-world simulation scenarios.

    \section{Conclusion}
    In this paper, we present a large-scale distributed parallel discrete event simulation engine for wargaming simulations, developed based on the Warped2 framework. 
    Our work introduces multiple internal and external enhancements to address the computational challenges of large-scale entity interactions. The experimental results demonstrate significant improvements in performance, scalability, and efficiency.
    This paper presents both internal and external improvements to enhance the framework's capabilities as follows.

    \begin{itemize}
        \item High-Throughput Processing with Listener Threads  
        
        To enhance the robustness and efficiency of the simulation engine, we incorporate listener threads, ensuring high-throughput operation while minimizing additional overhead. This improvement strengthens system integrity and contributes to the overall stability of large-scale simulations.
    
        \item Optimized Load Balancing via METIS Partitioning
          
        The METIS partitioning mechanism has been refined to achieve optimal load distribution across logical processes. Experimental results confirm that this approach effectively balances computational workloads, reducing bottlenecks and improving parallel efficiency.
    
        \item Large-scale Distributed PDES Engine Architecture Optimization \& Entity Search Algorithmic Acceleration  
        
        By reengineering the event communication structure and LP definition, we enable a scalable parallel simulation model. Additionally, the introduction of a nearest-neighbor search algorithm based on a single-level mesh reduces computational overhead by 34× compared to brute-force methods. These optimizations collectively enhance the solver’s efficiency and scalability for large-scale scenarios.
        
        \item Validated Performance \& Practical Scalability
        
        Through rigorous testing, our framework demonstrates an average 16× speedup over baseline implementations and 8× acceleration across diverse parallelization strategies. The Gridworld Demo further validates the engine’s applicability to real-world wargaming scenarios, confirming its capability to handle complex, entity-based simulations at scale.
    
    \end{itemize}

    Moving forward, we plan to explore further optimizations in event scheduling and parallel execution strategies to enhance performance. Additionally, we aim to extend the framework’s applicability to more complex multi-agent wargaming environments, incorporating adaptive partitioning techniques and AI-driven decision models to further improve simulation fidelity and efficiency.

    \section{Acknowledgment}
    This work was supported by the National Natural Science Foundation of China (Grant No. 12301560) and the Student Innovation Fund of Northwestern Polytechnical University (Taicang) (Grant No. TCCX240237).

\end{document}